\documentclass[prd, superscriptaddress, tightenlines, longbibliography, nofootinbib, eqsecnum, amsfonts, amsmath, floatfix, twocolumn, notitlepage]{revtex4-2}
\pdfoutput=1

\usepackage[utf8]{inputenc}
\usepackage{mathrsfs}
\usepackage{euscript}
\usepackage{graphics}
\usepackage{graphicx}
\usepackage{amsmath}
\usepackage{amssymb}
\usepackage{gensymb}
\usepackage{bm}
\usepackage[usenames,dvipsnames,svgnames,table]{xcolor}
\definecolor{linkcolor}{rgb}{0.6,0,0}
\definecolor{citecolor}{rgb}{0,0,0.75}
\definecolor{urlcolor}{rgb}{0.12,0.46,0.7}
\usepackage{xspace}
\usepackage{wasysym}
\usepackage{times}
\usepackage{appendix}
\usepackage{comment}
\usepackage{lipsum}
\usepackage[nolist,nohyperlinks]{acronym}
\usepackage{float}
\usepackage{simplewick}
\usepackage{natbib, ifthen}
\usepackage[breaklinks, colorlinks, urlcolor=urlcolor, linkcolor=linkcolor,citecolor=citecolor,pdfencoding=auto]{hyperref}
\usepackage{longtable}
\usepackage{listings}
\usepackage{colortbl} 
\usepackage{makecell}

\providecommand{\planck}{\textit{Planck}}
\providecommand{\Planck}{\planck}

\newcommand{\RDNzero}{\textrm{RD-}N^{(0)}}

\newcommand{\mksym}[1]{\ifmmode {\rm #1}\else #1\fi}

\newcommand{\taunl}{\ifmmode {\tau_{\rm{NL}}}\else $\tau_{\rm{NL}}$\fi}

\providecommand{\Planck}{\textit{Planck}}
\providecommand{\planck}{\Planck}

\providecommand{\text}[1]{\rm{#1}}

\providecommand{\arcmin}{{\rm arcmin}}

\providecommand{\Omm}{\Omega_{\mathrm{m}}}

\providecommand{\LCDM}{{$\rm{\Lambda CDM}$}}

\providecommand{\HEALpix}{{\tt HEALpix}}

\newcommand{\begm}{\begin{pmatrix}}
\newcommand{\enm}{\end{pmatrix}}

\newcommand\ba{\begin{eqnarray}}
\newcommand\ea{\end{eqnarray}}
\newcommand\bea{\begin{eqnarray}}
\newcommand\eea{\end{eqnarray}}

\newcommand\be{\begin{equation}}
\newcommand\ee{\end{equation}}

\defcitealias{Aghanim:2018oex}{PL2018}
\newcommand{\PL}{\citetalias{Aghanim:2018oex}\xspace}

\defcitealias{Akrami:2020bpw}{NPIPE2020}

\newcommand{\Simons}[0]{\emph{Simons Observatory}}

\newcommand{\eff}{\textrm{eff}}

\newcommand{\fid}{\rm fid}
\newcommand{\g}{g}
\newcommand{\R}{\mathcal{R}}

\newcommand{\MC}{\text{MC}}
\newcommand{\fpatch}{f_{A,L}}

\newcommand{\PRthree}[0]{PR3}
\newcommand{\PRfour}[0]{PR4}

\begin{document}
\title{CMB lensing from Planck \PRfour~maps}

\newcommand{\Sussex}{Department of Physics \& Astronomy, University of Sussex, Brighton BN1 9QH, UK}
\newcommand{\Geneve}{Universit\'e de Gen\`eve, D\'epartement de Physique Th\'eorique et CAP, 24 Quai Ansermet, CH-1211 Gen\`eve 4, Switzerland}

\author{Julien Carron}
\affiliation{\Geneve}
\affiliation{\Sussex}
\email{julien.carron@unige.ch}
\author{Mark Mirmelstein}
\affiliation{\Sussex}
\author{Antony Lewis}
\affiliation{\Sussex}

  \begin{abstract}
   {  	We reconstruct the Cosmic Microwave Background (CMB) lensing potential on the latest \planck~CMB \PRfour~(NPIPE) maps,
   which include slightly more data than the 2018 \PRthree~release, and implement quadratic estimators using more optimal filtering.  }
   {We increase the reconstruction signal to noise by almost $20\%$, constraining the amplitude of the CMB-marginalized lensing power spectrum in units of the \planck~2018 best-fit to $1.004 \pm 0.024$ ($68\%$ limits), which is the tightest constraint on the CMB lensing power spectrum to date. For a base \LCDM\ cosmology we find $\sigma_8 \Omm^{0.25} = 0.599\pm 0.016$ from CMB lensing alone in combination with weak priors and element abundance observations. Combination with baryon acoustic oscillation data gives tight 68\% constraints on individual \LCDM\ parameters $\sigma_8 = 0.814\pm 0.016$, $H_0 = 68.1^{+1.0}_{-1.1}$~km s$^{-1}$ Mpc$^{-1}$, $\Omm = 0.313^{+0.014}_{-0.016}$. \planck~polarized maps alone now constrain the lensing power to 7\%.}
  \end{abstract}


   \maketitle

\section{Introduction}

The \planck\ 2018 lensing analysis~\cite[hereafter PL2018]{Aghanim:2018oex} reconstructed the CMB lensing over most of the sky and derived powerful cosmological constraints, both on its own and in combination with other data. While ground-based experiments are rapidly improving the signal to noise of lensing reconstructions on small scales over parts of sky~\cite{Sherwin:2016tyf, SPT:2019fqo, Wu:2019hek, Darwish:2020fwf}, the \planck\ data is likely to remain the only full-sky dataset for the near future, and is the only data capable of reconstructing the largest-scale lensing modes. The analysis was sub-optimal in various respects, both due to the limitations of the processed data available at the time, and to various choices made to simplify the lensing analysis. In this paper we present a new analysis of the \planck\ data, with the aim of getting a more-optimal \planck\ lensing reconstruction over most of the sky. By using a different data processing pipeline we can also hope to get a clearer picture of how robust the \planck\ lensing constraints may be to unknown systematics and modelling uncertainties.

The latest \planck\ sky maps come from the NPIPE processing pipeline~\cite[hereafter NPIPE or \PRfour]{Planck:2020olo}, which uses ${\sim}8\%$ additional measurement time from the satellite repointing manoeuvres. NPIPE also improved the low-level data processing in many respects, and provides more simulations. In addition to using \PRfour~data, we also introduce and test a number of improvements to the \PL~analysis, gaining a scale-dependent improvement of up to ${\sim}20\%$ in signal to noise on the lensing power spectrum measurement. These include:
\begin{itemize}
	\item Joint inverse-variance (Wiener-)filtering of the temperature and polarization CMB maps, instead of the separate filtering performed in \PL. This uses the expected temperature to $E$-polarization cross-correlation to improve recovery of the $E$-mode map input to the quadratic estimator (QE). For \planck\ noise levels the joint lensing reconstruction is temperature dominated, and this translates into a modest ${\sim}3\%$ reduction of the Minimum Variance (MV) bandpower errors. The resulting lensing quadratic estimator is distinct from the optimized combinations of quadratic estimators presented in early work by Okamoto and Hu~\citep{Okamoto:2003zw, Hu:2001kj}, and was named GMV (Generalized Minimum Variance) by Ref.~\cite{Maniyar:2021msb}. The same reference shows that the improvement expected for more sensitive experiments is larger, and is expected to reach 10\% for \Simons~\cite{Ade:2018sbj,Maniyar:2021msb}.
	\item An additional post-processing step on the quadratic-estimated lensing convergence maps, by Wiener-filtering them according to the local reconstruction noise level on the sky. This step, introduced by Ref.~\cite{Mirmelstein:2019sxi}, has the virtue of weighting different parts of the sky more optimally when building the lensing power spectrum.
In the noise-dominated regime, the way the QEs are constructed automatically gives a nearly-optimal inverse-noise weighting~\citep{Aghanim:2018oex}. However, on scales where the signal is significant, a more uniform weighting is more optimal, and the extra filtering step gives an improved power spectrum estimate. In our case, this means a improvement of ${\sim}7\%$ in GMV bandpower errors close to the peak of the lensing spectrum at $L \sim 30$.
	\item Accounting for the noise inhomogeneity in our baseline GMV reconstruction by using an inhomogeneous noise variance in the CMB Wiener filter. As demonstrated by \PL~(but not used in the published likelihoods), this has a large impact on the quality of the polarization-only reconstruction, for which the CMB instrument noise dominates the polarization signal on small scales. Including the inhomogeneous filtering for the GMV estimator gives a total improvement of ${\sim}10\%$.
	\item Use of 600 end-to-end Monte-Carlo simulations (MCs), which is twice as many as was available for \PL. This allows us to reduce the errors in several terms calibrated with MCs (mean-field, MC- and RD-$N^{(0)}$-biases and covariance matrix).
	\item We use a novel way to obtain a realization-dependent covariance matrix, which takes better care of the inhomogeneities of the noise across the sky. This is discussed in Sec.~\ref{sec:covmat}.
\end{itemize}
\section{Analysis}
Sec.~\ref{sec:data} introduces the data sets and Sec.~\ref{sec:compsep} our foreground cleaning procedure. In Sec.~\ref{subsec:cmb_filtering} we discuss the filtering of the CMB maps and the construction of the quadratic estimators. Sec.~\ref{sec:kfilt} describes the additional filtering step we apply to the convergence map estimate. Sec.~\ref{sec:covmat} deals with our covariance matrix estimate, and Sec.~\ref{sec:PS} with the point-source correction we apply to the bandpowers. Finally, Sec.~\ref{sec:liks} discuss the construction of the lensing likelihood.

\subsection{Data and simulations}\label{sec:data}
We use NPIPE data maps for all frequency channels in three flavours (full mission maps, as well as the `A' and `B' splits designed to have maximally independent noise and systematics~\cite{Planck:2020olo}), together with the accompanying set of 600 end-to-end simulations of each map. The CMB sky signal simulation maps are the same as for \PRthree, and drawn from the \planck~FFP10 cosmological model\footnote{\url{https://github.com/carronj/plancklens/blob/master/plancklens/data/cls/FFP10_wdipole_params.ini}}. The noise components of \PRthree~and \PRfour~are however statistically independent.

A foreground component, based for each channel on the Commander sky model, is present in all simulations. On the main \planck~High Frequency Instrument CMB channels this contains resolved bright sources, galactic dust together with some level of large-scale CIB, and zodiacal emission. The data processing is intrinsically non-linear, so a perfect separation between the CMB, noise and foregrounds component of the simulations is not possible. The simulations that we are using are therefore not completely independent. However, as discussed in more detail in Sec.~\ref{sec:compsep}, the common component for the simulations is very small after foreground cleaning, and plays very little role in our analysis.

As in previous releases, there is a mismatch at the level of a few percent between the data and simulation power in temperature at the highest multipoles we use, $\ell \sim 2000$, presumably mainly due to residual foreground power. For this reason, before processing the maps to the lensing analysis, we add a simple Gaussian noise realization with a smooth power spectrum fit to the excess power in the foreground-cleaned simulations. This addition of power does not play an important role, except at the very low and even lensing multipoles where the mean-field is strongest (see Fig.~\ref{fig:verylowLs}   later on).

Besides using more data, the NPIPE processing introduced a number of modifications, or improvements, to the map-making process which are described in detail in Ref.~\cite{Planck:2020olo}. Given the complexity of the map-making process, it is difficult for us to single-out any one as being particularly critical to the lensing reconstruction. One major change in \PRfour~concerns large-scale polarization, where the use of template priors improves recovery of the polarized signal at large scales at the cost of a non-trivial transfer function. This is of no relevance to this paper, since low CMB multipoles do not carry much lensing signal at all, and we exclude the lowest CMB multipoles anyways.
\subsection{Component separation}\label{sec:compsep}
  \begin{figure}
   \centering
   \includegraphics[width=\columnwidth]{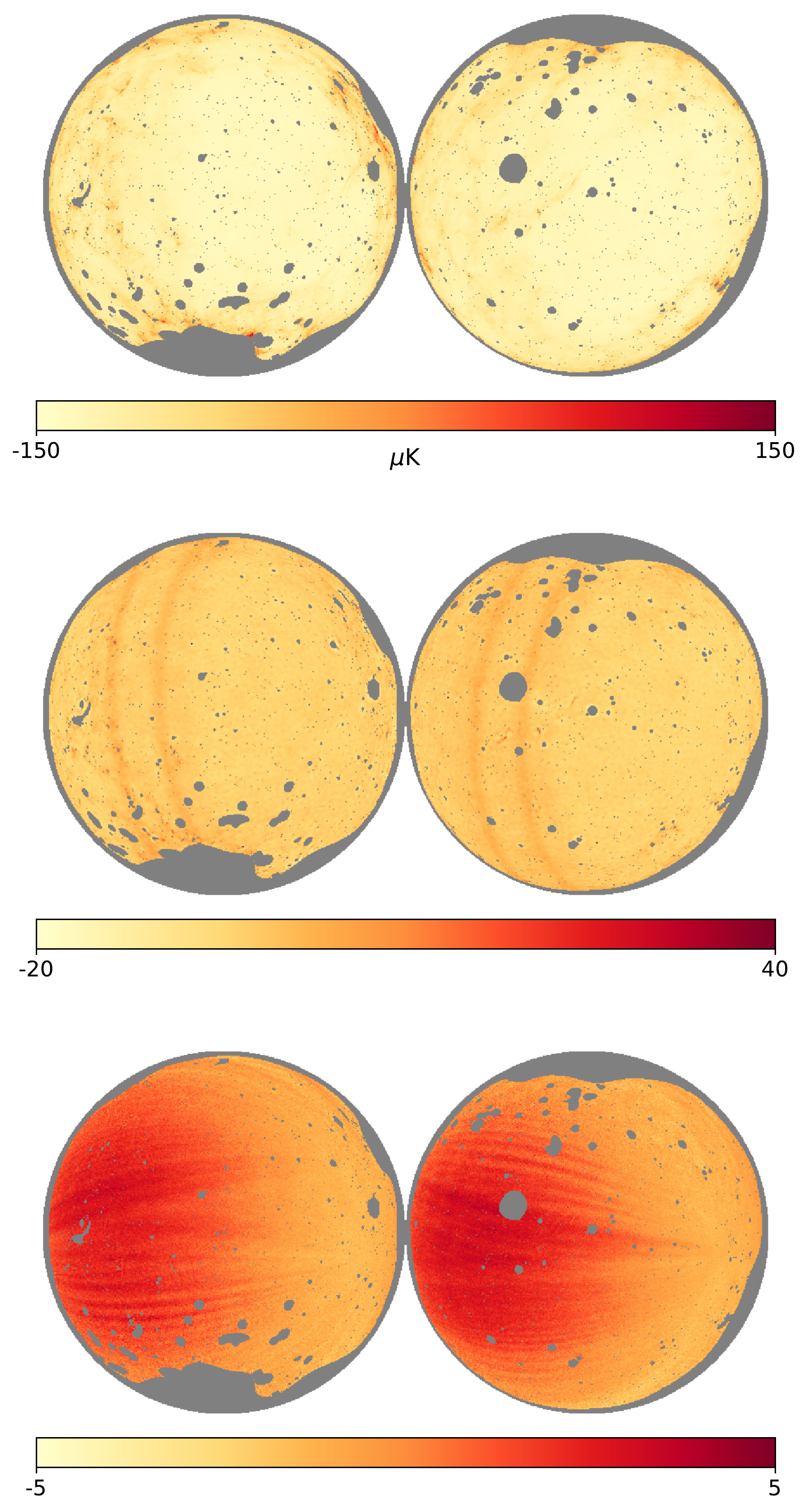}
   \caption{\label{fig:fgs}\emph{Top panel}: the foreground model map input to the cleanest CMB channel $143$GHz \PRfour~simulation maps on the lensing mask (all panels in $\mu$K, with different scales). \emph{Middle panel}: approximate residuals after foreground cleaning using our SMICA weighting of the frequency channels, where the most visible features are a handful of bright sources leaking slightly out of the point-source mask, and zodiacal emission dust bands. \emph{Bottom panel}: the difference of the middle panel to the \PRfour~simulations average, sourced by the data processing.
}
    \end{figure}
Previous \planck~releases provided foreground-cleaned CMB maps from four independent methods (SMICA, SEVEM, NILC and Commander), described in detail in Refs.~\cite{Planck:2015mis, Planck:2018yye} for the 2015 and 2018 release respectively. However, at the time of writing the only complete set of cleaned \PRfour\ temperature and polarization maps and simulations available were using the SEVEM component separation pipeline. For lensing reconstruction purposes from temperature, all four \planck~component separation methods perform comparably well. However, we found that SEVEM performs worse on the 2018 polarized data than the baseline SMICA~\cite{Cardoso:2008qt}, where the template fitting procedure of SEVEM~\cite{Fernandez-Cobos:2011mmt} is less effective than in temperature owing to the increased relative importance of the instrumental noise in polarization. For this reason, we built our own foreground-cleaned NPIPE maps, in the following way. SMICA linearly combines the frequency channels in harmonic space with a set of weights. Using the 2018 SMICA weights in temperature or polarization, with maximum multipole 2500, we simply rescaled them to account for the slight change in the effective transfer functions of each frequency channel, which we obtained empirically by cross-correlating the NPIPE frequency channel maps to the CMB FFP10 simulations input. Doing this could result in reduced suppression of foregrounds and slight suboptimality in the multipole range where both foregrounds and noise are relevant. However, we saw no evidence of a significant issue in the course of this analysis. This is not too surprising since \PL~showed that without any cleaning, expected biases from CIB contamination are at most 2\% on the CIB-bright 217GHz CMB channel without any cleaning, and down to ${\sim}0.1\%$ after SMICA weighting.

To build the \PRthree~temperature maps, the SMICA team applied an $f_{\rm sky}\sim 2\%$ preprocessing mask to the galactic centre and a few bright sources across the sky, and filled these pixels using diffusive inpainting prior to harmonic space weighting. If this mask is not included, the bright galactic centre produces ringing features that are distinctly visible when averaging simulations across the whole sky. We also suppress these pixels using an apodized version of the mask, but do not use an inpainting procedure on the \PRfour~channels. Our lensing analysis mask is then augmented to include the Boolean version of this apodized preprocessing mask. Most of the preprocessing mask is already present in the \PL~lensing mask, resulting in a negligible increase of 0.04\% in masked sky fraction.

The upper panel in Fig.~\ref{fig:fgs} shows the foreground template in the cleanest CMB channel simulations at $143\rm{GHz}$. Our SMICA weighting of these input foregrounds, shown on the middle panel, gives a crude impression of the cleaning efficiency. Aside from a couple of zodiacal emission dust bands~\cite{Planck:2013maj}, the most visible residuals are a handful of sources of intensity a few tens of $\mu \rm{K}$, which are the point-source component leaking a bit outside of the analysis mask. The point source mask was built for the 2015 \Planck~lensing analysis~\cite{Planck:2015mym} based on the \planck~catalogue of sources at 143 and 217$\rm{GHz}$, and we find that extending the mask slightly would remove these residuals. However, they have no impact on the reconstruction (see Fig.~\ref{fig:TTvariations} later on). The weighted foreground residual template varies, but only at the level of a few $\mu \rm{K}$, from the simulated data-processed one as shown on the lower panel. This was obtained by taking the average of the 600 \PRfour~simulations, after subtracting from each simulation an approximation to their data-processed CMB component, where the isotropic effective \planck~\texttt{QuickPol} beams\cite{Hivon:2016qyw,Planck:2019nip} were applied to the input sky.
 From the SMICA weights, and the pixel covariance maps produced by the NPIPE processing for each channel, we also constructed the pixel variance maps that we use in the Wiener-filtering of the combined CMB maps. We do so assuming white and independent noise across channels, and independent Stokes T, Q and U noise, following \PL[Eqs.~(15-17)]. We then rescale these maps with a constant factor such that in harmonic space this white noise value matches  on the relevant (small scales) CMB multipoles approximately the empirical noise power.

\subsection{CMB filtering and quadratic estimators}
\label{subsec:cmb_filtering}
Given the Stokes $T$ and $_{\pm 2}P \equiv Q\pm iU$ data provided on the pixelized sky, the maximum a posteriori (Wiener-filtered) modes are given by
	\begin{equation}\label{eq:filt}
		\begin{pmatrix}
		T^{\rm WF} \\ E^{\rm WF} \\ B^{\rm WF}	
		\end{pmatrix} \equiv C^{\rm fid} \mathcal T^\dagger \left[\mathcal T C^{\rm fid} \mathcal T^\dagger + N\right]^{-1} 		\begin{pmatrix} T^{\rm dat} \\ Q^{\rm dat} \\ U^{\rm dat} \end{pmatrix},
	\end{equation}
where $\mathcal T$ is our fiducial transfer function, that maps lensed $T, E, B$ harmonic modes onto the observed Stokes maps (inclusive of the 5~\arcmin~beam and the \HEALpix\ pixel window function). $C^{\rm fid}$ is a matrix of our fiducial lensed CMB spectra, with elements
\begin{equation}
 \left[C^{\rm fid}\right]^{XY}_{\ell m, \ell'm'}	= \delta_{\ell \ell'}\delta_{mm'}\begin{pmatrix}
		C_\ell^{ TT,\rm fid} &C_\ell^{TE, \rm fid} & 0 \\
		C_\ell^{TE,\rm fid} &C_\ell^{EE, \rm fid} & 0\\
		0 &0 & C_\ell^{BB,\rm fid}
	\end{pmatrix},
\end{equation}
and $N$ is the noise matrix for the real-space Stokes data, which we assume independent between pixels, and is constructed as described in section~\ref{sec:compsep}. The fiducial spectra are those of the FFP10 cosmology, the same for the \PRfour~and \PRthree~simulation sets. The filtered maps are calculated using the conjugate-gradient method following \PL~\cite{Smith:2007rg}, with an improved convergence criterion putting more weight on the CMB sub-degree scales relevant for the lensing signal, resulting in faster filtering by almost a factor of two.

Foreground templates, or any set of poorly-understood modes, can be removed from the analysis by augmenting the noise matrix of Eq.~\eqref{eq:filt} and assigning them an infinite noise value, a procedure sometimes called `deprojection'. For example, we always deproject the temperature monopole and dipole. We also tested deprojection of the SMICA-weighted Commander foreground model, giving perfectly consistent results as shown later on.

Using these filtered CMB maps, quadratic estimators (denoted $\hat g_{LM}^{\phi}$) are built in the same way as \PL[Eqs.~(3-6)]. As there, we filter CMB multipoles $2\leq \ell \leq 2048$ but only use filtered multipoles $100 \leq \ell \leq 2048$ to construct the QEs.

\subsection{$\kappa$-filtering and lensing spectrum}
	\label{sec:kfilt}
\begin{figure}
\centering
   \includegraphics[width=0.8\columnwidth]{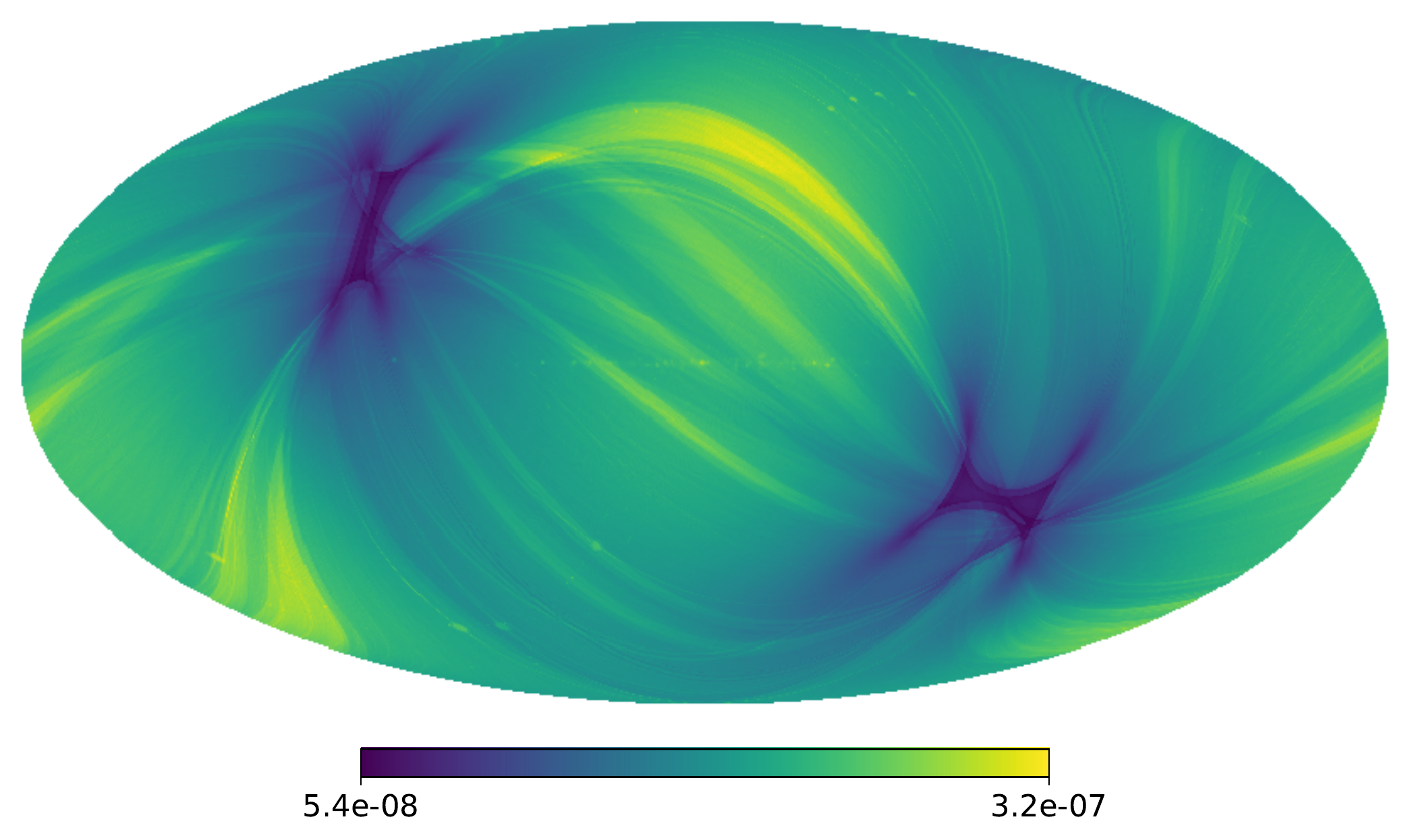}
      \caption{The large-scale effective lensing convergence reconstruction noise variance map, $N^\kappa_{0,{\rm eff}}$, obtained by averaging the approximately flat convergence noise spectrum across the range $8 \leq L \leq 100$.
      After applying the same mask as for CMB filtering, and condensing this map into a simpler one with 64 homogeneous noise patches, this map is used to perform the filtering of the quadratic estimator convergence maps that are then used for our Minimum Variance estimate.
      This $\kappa$-filtering operation down-weights regions where the lensing reconstruction is poorest (the ecliptic equator), decreasing the bandpower errors. The overall gain remains limited, since it is CMB fluctuations, which are isotropically distributed across the sky, that dominate the reconstruction noise rather than instrumental noise.}
         \label{fig:N0map}
   \end{figure}

The scanning strategy of the \planck~satellite results in many more scans of the ecliptic poles compared to the equator, resulting in inhomogeneous noise across the sky. This implies that the expected lensing reconstruction noise $N^{(0)}$ also varies depending on the sky location. The expected GMV large-scale lensing convergence reconstruction noise is shown in Fig.~\ref{fig:N0map}, spanning almost one order of magnitude.

In order to account for this, once the QEs are obtained, we filter them following the methodology which was introduced by Ref.~\cite{Mirmelstein:2019sxi}. The unmasked regions of the sky are partitioned into 64 patches (based on their relative pixel CMB temperature noise level). This partitioning is done by building a histogram of the inverse noise map. The temperature and polarization noise variances in each patch (histogram bin) are then approximated as homogeneous, which is the main requirement for this filtering step. This partitioning scheme is used to construct the effective lensing reconstruction noise level of each patch, $N^\kappa_{0,\eff}$, which is used in the filtering process. The local patch reconstruction noise values are calculated from averaging the values of $N_{0,L}^{\kappa}$ (the isotropic reconstruction noise $N_{0,L}$ for $\kappa_{LM} \equiv L(L+1) \phi_{LM} / 2$) over the multipole range $8\leq L\leq 100$. In this multipole range the reconstruction noise is relatively white. The same partitioning scheme is also used to construct an effective response map, $\R_{\eff}^{\kappa}$, which is used to properly normalize the QEs. The values for the response map $\R_{\eff}^{\kappa}$ are obtained in a similar way from the response $\R_{L}^{\kappa}$. Fig.~\ref{fig:N0map} shows the reconstruction noise map which is used in the filtering. We set $N_{0,\eff}^{\kappa}(\hat n)^{-1}=0$ for all masked pixels. The additional filtering step is performed on the quadratic estimator convergence $\kappa$, as it is approximately local in real space, and hence is uncorrelated between patches on large scales and has approximately white noise.

Similarly to Eq.~\eqref{eq:filt}, we define the Wiener-filtered estimator as
\begin{equation}
\hat{\kappa}^{\rm WF} \equiv C^{\kappa\kappa}_{\fid}\left[C^{\kappa\kappa}_{\fid}+N^\kappa_{0,\eff}\right]^{-1} \left(\R_{\eff}^{\kappa} \right)^{-1} \hat \g^{\kappa},
\label{eq:kappa_filt}
\end{equation}
where $C_{\fid}^{\kappa\kappa}$ is the fiducial theory $\kappa$ power spectrum, and the QE map $\hat{\g}^{\kappa}$ is defined after mean field subtraction from\footnote{The relation \eqref{eq:kmf} is inverse to that of the lensing convergence $\kappa$ to the potential $\phi$ because the unnormalized quadratic estimator $\hat g$ are gradients with respect to these quantities.}
\begin{equation}\label{eq:kmf}
\frac 12 L(L + 1) \hat{\g}_{LM}^{\kappa} \equiv \left(\hat{\g}_{LM}^{\phi}-{\langle\hat{\g}_{LM}^{\phi}\rangle}_{\MC}\right).
\end{equation}
The bracketed matrix inverse in Eq.~\eqref{eq:kappa_filt} is the only non-trivial step, since the first and second terms in that matrix are diagonal in harmonic and pixel space respectively. This step is performed using the same conjugate gradient method and preconditioner as used for filtering the CMB maps (Sec.~\ref{subsec:cmb_filtering}), which results in fast enough convergence to the solution. Eq.~\eqref{eq:kappa_filt}, the `$\kappa$-filtered estimator', gives us a novel (yet unnormalized) lensing map whose spectrum is by construction approximately optimally weighted across the sky. We produce these maps in pairs $\hat \kappa_1, \hat \kappa_2$, with subtracted mean-field estimates in Eq.~\eqref{eq:kmf} built from two non-overlapping subsets of simulations. This avoids the mean-field Monte-Carlo noise in the raw spectrum estimate, given by
\begin{eqnarray}
C_L^{\hat\kappa\hat\kappa} \equiv \frac{1}{(2L+1)f_{A, L}} \sum_{M = -L}^L\hat \kappa^{\rm WF}_{1, LM}\hat \kappa^{\rm WF, *}_{2, LM}.
\label{eq:clkappa_biased}
\end{eqnarray}
The factor $f_{A,L}$, a scale-dependent effective sky fraction, encapsulates the effects of masking and of the $\kappa$-filtering normalization. Using the independent patch approximation, we may write~\cite{Mirmelstein:2019sxi}

\begin{eqnarray}
\fpatch \equiv \sum_{\textrm{patches } p} f_{p} \left(w_L^{p}\right)^2
\label{eq:f_A}
\end{eqnarray}
where $f_p$ is the sky area of patch $p$, and
\begin{eqnarray}
w^{p}_L=\frac{C_{L,{\fid}}^{\kappa\kappa}}{C_{L,{\fid}}^{\kappa\kappa} + N^{\kappa,p}_{0,{\eff}}} \frac{\R^{\kappa,p}_L}{\R^{\kappa,p}_\eff}.
\label{eq:wkappa}
\end{eqnarray}
Beside the lensing spectrum signal, the raw spectrum in Eq.~\eqref{eq:clkappa_biased} still contains a number of additional terms, the most relevant for \planck~noise levels being the two of lowest orders in the lensing potential $N_L^{(0)}$ and $N_L^{(1)}$~\cite{Hanson:2010rp, Kesden:2003cc}. We use the now very standard realization-dependent RD-$\hat N_L^{(0)}$~\cite{Planck:2015mym, Story:2014hni} and simulation-based MC-$N_L^{(1)}$~\cite{Story:2014hni, Aghanim:2018oex} debiasers, which can be applied without modifications to the GMV and $\kappa$-filtered maps. At higher order, $N_L^{(3/2)}$ induced by the large-scale structure bispectrum and post-Born lensing is negligible at \planck~noise levels~\cite{Bohm:2016gzt, Bohm:2018omn, Beck:2018wud, Fabbian:2019tik, Pratten:2016dsm}, and $N_L^{(2)}$ is made negligible using the lensed CMB spectra in Eq.~\eqref{eq:filt}~\cite{Hanson:2010rp}.

After bias subtraction, we apply a multiplicative Monte-Carlo correction to the bandpowers just as in \PL, by comparing simulated reconstruction power spectra with input lensing power. In previous analyses, this residual Monte-Carlo correction is mostly sourced by the presence of the mask: masking renders the analytic isotropic normalization of the QE inaccurate close to the mask boundaries, while application of a correct anisotropic normalization appears intractable. In our case, both inhomogeneous CMB- and $\kappa$-filtering impacts the response of the QE to lensing, hence its normalization as well. Expression~\eqref{eq:wkappa} and the isotropic rescaling of Eq.~\eqref{eq:clkappa_biased}, which originates from the approximate patch model, is of course not exact. However, all terms of the debiased spectrum scale according to the same $1/f_{A,L}$ so that a modelling error is captured automatically by the Monte-Carlo correction. The true response function only depends on the sky signal CMB spectra, which are very well constrained and whose variations always play a very subdominant role in \planck~lensing likelihoods. As long as the correction is small, its cosmology dependence may be completely neglected. In previous releases it was at most ${\sim} 10\%$ on the lowest multipole bins. With the $\kappa$ filtering we now find that the residual MC correction is slightly smaller (which must be seen as a coincidence rather than a virtue of the model).

We finally apply a (very small) point-source (PS) correction $ \Delta \hat C_L^{PS}$, discussed in more detail in Sec.~\ref{sec:PS}. Our final bandpowers may then be written as
\begin{eqnarray}
\hat C_{L}^{\kappa\kappa} \equiv C_L^{\hat\kappa\hat\kappa}-\textrm{RD-}\hat N_L^{(0)} - \textrm{MC-}\hat N_L^{(1)} - \Delta \hat C_L^{PS}.
\label{eq:clkappa}
\end{eqnarray}

We find that with $\kappa$ filtering, the reconstructed bandpower variance is slightly improved around the peak of the reconstructed power spectra; not only for our GMV reconstruction (see Fig.~\ref{fig:sigmas}), but also for the polarization-only reconstruction, which is much more noise-dominated. This is quantified in Sec.~\ref{sec:polonly}. For more sensitive data from future SO and CMB-S4 experiments, the expected improvement was already demonstrated using a fiducial small flat-sky area scanning strategy in Ref.~\cite{Mirmelstein:2019sxi}. Revisiting their analysis for realistic curved-sky scanning strategies from Chile similar to those of Ref.~\cite{Stevens:2018biw} we find similar results, with variance improvements in the range between 5-25\%. This improvement clearly depends primarily on the homogeneity level of the scan; there is greater improvement for inhomogeneous scans.

\subsection{Covariance matrix}\label{sec:covmat}
\newcommand{\NRD}{\ensuremath{N_{\rm RD}}}
The data bandpowers are debiased with the usual realization-dependent $\textrm{RD-}\hat N_L^{(0)}$ estimate, which uses quadratic estimators that cross maps from the data with simulations. While this improves the covariance of the estimator, it is too expensive to perform the same type of realization-dependent subtraction on each and every simulation. There are at least a couple of options available:
\begin{itemize}
	
\item One option is to consider the covariance matrix obtained without (or, with realization-\emph{independent}) debiasing. This slightly overestimates  both the variance and covariance of the binned bandpowers, particularly so on small scales~\cite{Peloton:2016kbw}.
\item Another possibility is to use a numerically cheap approximate realization-dependent debiaser: for example, \PL~used the analytical formula for the $N_L^{(0)}$ noise on an isotropic sky with homogeneous filtering, but replacing the idealized power spectra by the empirical ones, to obtain a semi-analytic $\hat N_L^{(0)}$ estimate. The accuracy of this estimate is insufficient for the purpose of precision debiasing on realistic data. It is, however, quite efficient at incorporating the removal of the small-scale covariance, and, using homogeneous CMB filtering, appears good enough for a covariance matrix estimate.
 \end{itemize}
Here we try to improve on the second option, by building a cheap debiaser from simulations which takes into account the filtering used on the data, and makes no assumptions of isotropy. We first discuss the case without $\kappa$-filtering:
For each of 480 simulations dedicated to the covariance matrix estimate, we build an $\textrm{RD-}\hat N_L^{(0)}$ estimate using a much smaller number (\NRD) of other simulations than are used to debias the data. These simulations come from the same original set, requiring the filtering of no additional maps, and we neglect the impact on the covariance of simulation subset overlap. The additional QE's ($\NRD \times 480$) that are required are built on the fly from the filtered maps. Since $\NRD$ is small, the reconstruction noise part of the covariance is overestimated slightly. The proportionality constant is $4 / N_{\rm RD}$ as derived by \PL, App. C. This motivates the following procedure:
for a given \NRD, we build a correction $c^{\NRD}_L$ to the empirical full covariance matrix $\hat \Sigma^{\rm RD}$ and rescale it as
\begin{equation}
	\hat \Sigma_{LL'} \equiv \frac{\hat \Sigma^{\rm RD}_{LL'}}{\sqrt{c^{\NRD}_L c^{\NRD}_{L'} }}
\end{equation}
with
\begin{equation}
	c_L \equiv 1 + \frac{4}{\NRD} \left( \frac{ \textrm{MC-}\hat N_L^{(0)}}{C_L^{\hat \phi \hat \phi}} \right)^2,
\end{equation}
where $C_L^{\hat \phi \hat \phi}$ is the mean-field subtracted raw data power. We find that even for very small \NRD~this estimate is quite stable as a function of \NRD, and we used $N_{\rm RD} \sim 16$ for our final estimates. A value of $N_{\rm RD}$ that is too large is both costly and not desirable: with all the simulations coming from the same set, the overlap would reintroduce the correlations that the realization-dependent debiasing removes.
This covariance shares the characteristic qualitative features expected for $\RDNzero$-debiased estimate;  that is, less covariance between the spectrum estimates at different lensing multipoles, as well as lower variance at the highest lensing multipoles. With homogeneous filtering, this estimate is also numerically closer to the semi-analytic debiasing result of \PL~than without $\rm{RD}$-debiasing. On the conservative range the size of the effect is visible but remains a few percent.

 The $\kappa$-filtering operation of Sec.~\ref{sec:kfilt} has a small effect close to the peak of the lensing power. To build a realization-dependent covariance without having to $\kappa$-filter a large number of cross-simulation maps, we simply rescale the non-$\kappa$-filtered covariance in the following manner: with $r_{L_b}$ the ratio between the realization-independent variances in each bin $b$ with and without $\kappa$ filtering, we multiply the realization-dependent, non $\kappa$-filtered $b,b'$ covariance by $\sqrt{r_{L_b} r_{L_{b'}}}$.

\begin{figure}
   \centering
   \includegraphics[width=\hsize]{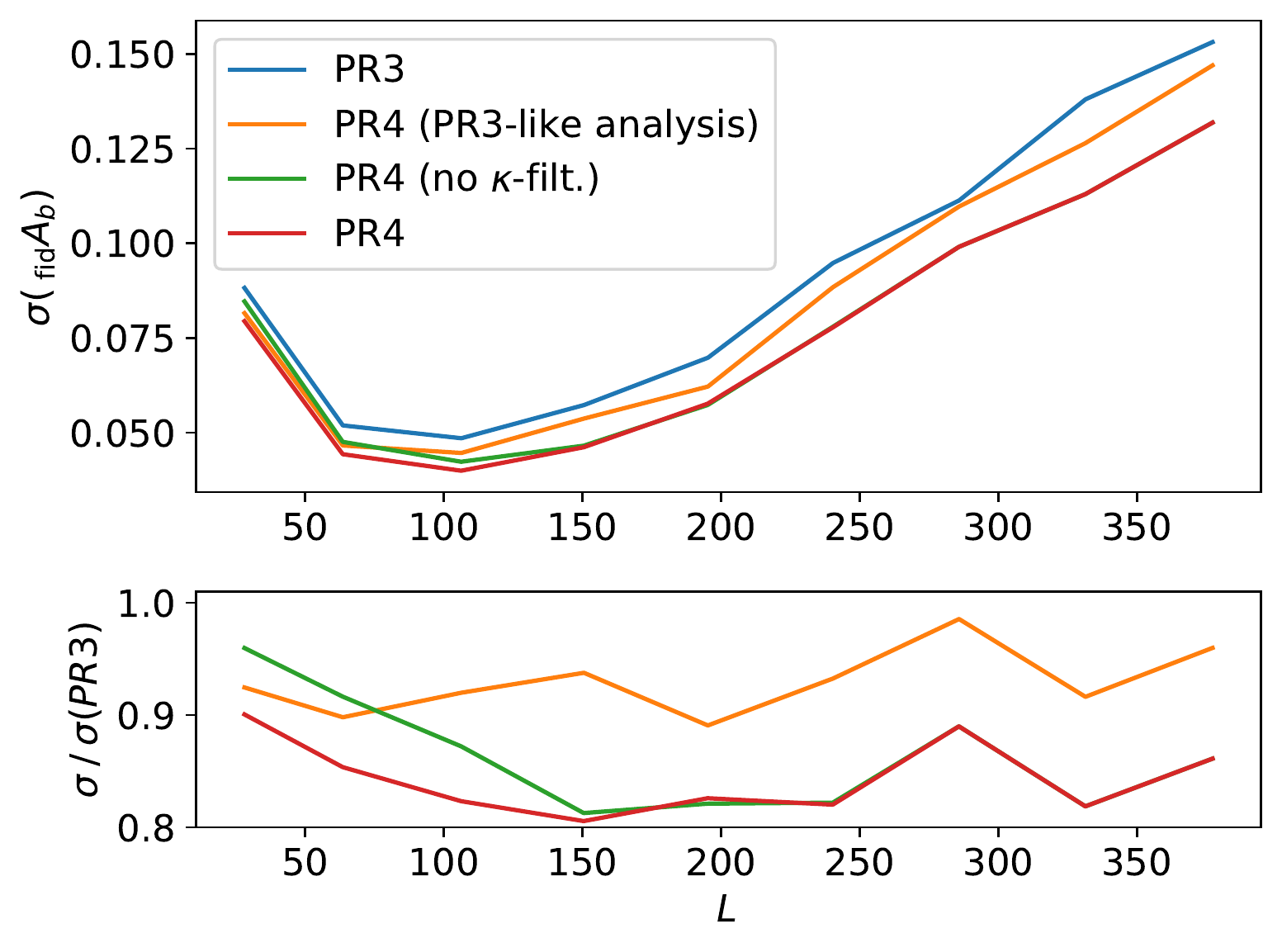}
      \caption{\emph{Top panel}: 
      Error bars on our reconstructed lensing power spectrum bandpower amplitudes $_{\rm fid}A$
      relative to the FFP10 fiducial model, over the conservative range $8 \leq L \leq 400$.
      The inhomogeneously CMB-filtered and $\kappa$-filtered GMV bandpowers, with $\kappa$-filtering (our best result, red), 
      and without $\kappa$ filtering (green), are shown compared to the published MV \PL reconstruction (blue). The orange line shows the results of applying the same 2018 pipeline to the NPIPE maps.
      \emph{Bottom panel}: The ratio of these errors to those of the \PL~reconstruction.
      The reconstruction without $\kappa$-filtering, but using improved CMB-filtering with respect to \PRthree\ (green), performs worse at low multipoles than the \PRthree-like analysis in agreement with purely analytic predictions.
      This is because using inverse reconstruction noise $N^{(0)}$ weighting for power spectrum estimation, as implied without $\kappa$-filtering, is suboptimal on large scales where the first couple of bins are now highly signal dominated. This is corrected by $\kappa$-filtering, which produces a reconstruction with (very) slightly higher $N^{(0)}$ bias but smaller error bars by effectively using more of the sky.
      }
         \label{fig:sigmas}
   \end{figure}

The top panel of Fig.~\ref{fig:sigmas} shows the error bars on our most precise reconstruction (red), together with variations (orange and green) and the \PRthree~baseline result (blue). The bottom panel shows the corresponding ratio to the \PRthree~errors. For most bins the improvement that we see in this work is about half from the slightly reduced noise of \PRfour, and half from lensing pipeline improvements.

\subsection{Point-source correction}\label{sec:PS}
It is well known that lensing estimators can respond strongly to point-source-like signals in the CMB data~\cite{Osborne:2013nna,Ade:2013tyw}. To check for a residual point-source signal, we apply the point-source anisotropy estimator $S_2(\hat n)$ to our maps. After subtraction of the expected lensing-induced trispectrum contamination to the $\hat S_2$ spectrum, we obtain a result consistent with a white noise spectrum, and non-zero at about $3$-$4\sigma$. We compress this spectrum further into a trispectrum shot-noise amplitude $S_4$. The details of this step differ slightly from the $\Planck$ 2015 and 2018 analyses, since we noticed that on simulations our $\hat S_2$ spectrum estimates were highly correlated on most scales\footnote{This large covariance is without a realization-dependent debiaser, which could help further improve the estimate of the point-source signature, but at substantial numerical cost.}, with a positive cross-correlation greater than 50\% at $L > 500$. Taking this into account shifts the weighting of the $\hat S_2$ spectrum towards larger scales, and provides us with a more precise estimate, reducing the error on $S_4$ by 30\% with respect to the \planck~analysis weighting. We measure
\begin{equation}
	10^{13}\:\hat S_4 = 3.0 \pm 0.8 \: \mu\rm{K}^4,
\end{equation}
an estimate larger by about $30\%$ than the result from the \PRthree~maps.
This non-zero trispectrum affects our $\kappa$-filtered lensing power according to
\begin{equation}
	f_{A, L}\:\Delta \hat C_L^{PS} = \hat S_4 \sum_p f_p \left(w_L^{p}\right)^2 \left(\frac{\mathcal R_L^{\kappa s, p}}{\mathcal R^{\kappa, p}_{L}}\right)^2
\end{equation}
where $\mathcal R^{\kappa s, p}_L$ is the local response of the lensing estimator to the point source field. The effect of the inhomogeneous filtering is small, the correction having very much the same shape as for more standard homogeneous filtering where all terms in the sum are identical.
This correction reaches a fourth of the bandpower error on the last few bins of the conservative range, and is smaller elsewhere. To account for the uncertainty in $S_4$, the covariance matrix is augmented by the rank-one $\sigma^2_{S_4} \mathcal P \mathcal S\: \mathcal \cdot \mathcal P\mathcal S^t$, where
\begin{equation}
	\mathcal P\mathcal S_L \equiv \frac {1}{f_{A, L}}\sum_p f_p \left(w_L^{p}\right)^2 \left(\frac{\mathcal R_L^{\kappa s, p}}{\mathcal R^{\kappa, p}_L}\right)^2
\end{equation}
 is the point-source correction template (before binning and application of the final percent-level Monte-Carlo correction) and $\sigma^2_{S_4}$ is the squared error on the trispectrum quoted above. Added in quadrature to the covariance, this is everywhere a very tiny correction to the bandpower errors: at most 0.3\% on the conservative range and practically zero on the lensing power peak, but positively correlated across the lensing potential spectrum tail.

\subsection{Likelihoods}\label{sec:liks}
We produce likelihoods for our improved bandpowers, generalizing the approach of \PL. The construction of the likelihoods for the CMB lensing bandpowers is complicated by the fact that, as well as direct dependence on $C^{\phi \phi}_L$ and via $N^{(1)}$, the estimators also carry dependence on cosmology through the lensed CMB power spectra entering $N^{(0)}$, $N^{(1)}$, the mean-field and the estimator normalization. We follow previous approaches, making use of the fact that the CMB spectra are very well constrained, which can be summarized as
\begin{itemize}
	\item neglecting the cosmology dependence of the RD-$N^{(0)}$ debiasing (since by construction it is insensitive to leading order differences between the fiducial and true sky CMB spectra),
	\item linearizing the response functions and $N^{(1)}$ matrix to small changes in the CMB spectra,
	\item neglecting the cosmology dependence of the residual MC and PS corrections (which are small),
	\item and neglecting the cosmology dependence of the mean field. The dominant source of mean field is masking, which creates a strong signature directly proportional to its isotropic power spectrum even for perfectly Gaussian data. In our case, the mean field is still sizeable on the first bin of the conservative range. Unless the simulation power has been matched accurately to the measured data power the mean field sourced by the CMB signal has a cosmology dependence. By looking at the data splits this can be inferred to be between a third and half of the total. However, since the mean field is subtracted at the map level (and not from the bandpowers), the impact is quadratic in the mismatch of CMB spectra, hence completely negligible except at the very lowest multipoles (see Fig.~\ref{fig:verylowLs}, brown points).

\end{itemize}
This allows us to produce a fast and accurate likelihood for most cases of practical interest.
In our case there is one difference to previous work: since we use inhomogeneous filtering, the responses and bias terms vary according to the location on the sky. We take this into account by approximating the sky as consisting of patches of homogeneous noise levels, so that the full linear response matrix is itself the sum of the same number of such matrices, each built for a different filtering noise level. As discussed in section~\ref{sec:kfilt} this approximation is very effective and gives a residual MC-correction (sourced predominantly by the presence of the analysis mask) of the same size at the more standard case of homogeneous filtering (where there is a single response function and $N^{(1)}$ across the sky).

Explicitly, our likelihood is constructed as follows. Considering the decomposition of the sky in patches with roughly homogeneous filtering noise levels, we can consider that the lensing map estimate in patch $p$ responds to the sky signal as
\begin{equation}
 	\hat \kappa^{\textrm{WF}, p}_{LM}(\theta) \sim w^p_L \frac{\mathcal R_L^p(\theta)}{\mathcal R^{p}_L(\theta^{\rm fid})} \kappa^{\rm sky}_{LM}.
\end{equation}
For homogeneous CMB- and no $\kappa$-filtering there is a single patch, with unit $w^p_L$, and we recover the \PL~likelihood construction. In this more general case we can write the linearized likelihood prediction for the observed spectrum prior to binning
\begin{equation}\label{eq:lik}
\begin{split}
	C_L^{\kappa \kappa, \rm pred}(\theta) &= C_L^{\kappa\kappa}(\theta) +	\sum_{L'}\mathcal M^{\kappa \kappa}_{LL'}\left(C^{\kappa \kappa}_{L'}(\theta)-C^{\kappa \kappa}_{L'}(\theta^{\rm fid}) \right) \\
	&+ C_L^{\kappa \kappa}(\theta^{\rm fid})\sum_{\ell, XY} \mathcal M^{XY}_{L\ell}\left( C_\ell^{XY}(\theta) -  C_\ell^{XY}(\theta^{\rm fid})  \right)
\end{split}
\end{equation}
where the matrices involved are
\begin{equation}
	\mathcal M^{XY}_{L \ell} = \!\!\!\!\sum_{\textrm{patches } p} \!\!f_p \left(\tilde w_L^{p}\right)^2 \left(\frac{\partial \ln \left(\mathcal R_L^{p}\right)^2(\theta^{\rm fid})}{\partial C_\ell^{XY}} +\frac{\partial N_L^{(1), p}(\theta^{\rm fid})}{\partial C_\ell^{XY}} \right)
\end{equation}
for $XY \in (TT, TE, EE)$, and
\begin{equation}
	\mathcal M^{\kappa \kappa}_{LL'} = \sum_{\textrm{patches }p}f_p \left(\tilde w_L^{p}\right)^2 N^{(1), p}_{LL'}(\theta^{\rm fid})
\end{equation}
with $\tilde w_L^p \equiv w_L^p / f_{A,L}$.
In order to get lensing-only constraints, we make the assumption that the fluctuations in the CMB spectra in the second line of Eq.~\eqref{eq:lik} are Gaussian. We can then analytically marginalize these out, obtaining a likelihood that depends only on the cosmology's lensing power. For this we proceed just as \PL, using the smoothed \PRthree~\texttt{plik\_lite} CMB bandpowers and covariance matrix. This leads to an increase of the lensing covariance by several percent, and to a small shift in power sourced by the small differences between the FFP10 and the \planck~empirical spectra. We also produce a likelihood for the polarization-only reconstruction. It is of course much noisier, and for this likelihood we neglect the CMB spectra corrections altogether.

Construction of the likelihood requires calculations of many $N^{(1)}$ lensing matrices and their derivative with respect to the CMB spectra. To perform this in a more efficient way than in previous releases, especially for the GMV estimator, we have written a novel fast Fourier transform based algorithm for their calculation\footnote{\url{https://github.com/NextGenCMB/lensitbiases}}, in many cases orders of magnitude faster than previous methods.

\section{Results}
  \begin{figure*}
   \centering
   \includegraphics[width=\textwidth]{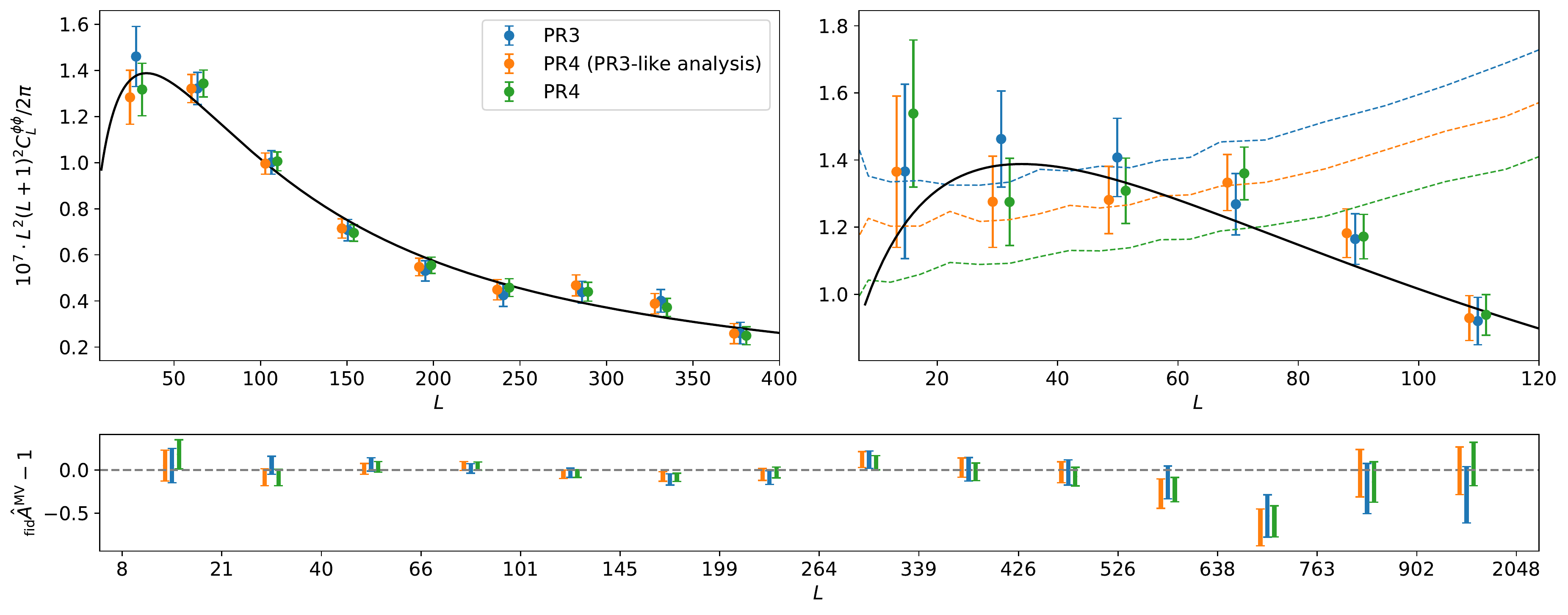}
   \caption{The left panel shows our most precise bandpower estimates over the conservative multipole range (green), together with the \planck~2018 (\PRthree) Minimum Variance bandpowers (blue). The black line is the lensing power spectrum best-fit to the \planck~2018 CMB spectra data (without lensing). The right panel focuses on lensing multipoles close to the $\Lambda$CDM lensing power spectrum peak, and also shows as the dotted lines the corresponding realization-dependent $N^{(0)}$ biases (without residual Monte-Carlo correction). Many more modes are now signal dominated. All bandpowers are built with the same approximate inverse-variance multipole weighting scheme given in \PL. The orange points show the results of implementing the \PRthree~reconstruction analysis on the NPIPE maps by using a homogeneous noise sky model as well as independent temperature to polarization filtering. The bottom panel shows the relative deviation to the fiducial FFP10 cosmology amplitude on the entire range.\label{fig:cppcomp22018}}
    \end{figure*}
\subsection{GMV and TT reconstructions}\label{sec:GMV}
Fig.~\ref{fig:cppcomp22018} shows our bandpower reconstructions on the conservative range $8 \leq L \leq 400$ (green), together with the \PL~results (blue).
Table~\ref{table:tableMV} lists our GMV, $\kappa$-filtered bandpower values which we later use to constrain parameters.
To highlight the differences caused by the new \PRfour~data processing, Fig.~\ref{fig:cppcomp22018} also shows our results on the new NPIPE maps with the same lensing analysis choices as 2018 (orange). The same binning scheme is used in all three cases. The reconstructions do display some differences: the right panel provides a closer view around the lensing spectrum peak, with a finer binning, where several shifts comparable to the error bars can be seen. Summary statistics and broad characteristics of the spectrum remain very similar however. Using our CMB-marginalized likelihood (see Sec.~\ref{sec:liks}), we fit an amplitude of the lensing spectrum relative to the $\Lambda$CDM prediction from the best-fit to the 2018 \planck~CMB spectra (without lensing, \texttt{plikHM\_TTTEEE\_lowl\_lowE})~\cite{Planck:2018vyg}. On the conservative range we get
\begin{equation}
	_{\rm b.f.}A^{\rm GMV}_{8 \rightarrow 400} = 1.00 \pm 0.024,
\end{equation}
with a $\chi^2$ of $8.5$ for $9$ data points, with 2018 value $1.01 \pm 0.026$\footnote{The shift in amplitude is closer to 0.003 before rounding.}. Both spectra show the same slight tilt with respect to the best fit, with slightly higher (lower) amplitude preferred at low (high) multipoles. Owing to the shift on the first conservative bin this tilt is only slightly less pronounced for \PRfour. This property is slightly more pronounced for the temperature-only reconstruction, where the two lowest conservative bins are higher than their \PRfour~counterparts by approximately $1\sigma$ using the same lensing pipeline.

Given the complexity of the data processing, `paired' noise simulations --- sharing the same random seeds between the 2018 release and NPIPE --- are neither available nor planned. This makes it difficult to evaluate if the peak shifts seen can be explained solely by the data processing differences. We attempted to quantify this in the following approximate manner\footnote{We thank A. Challinor for suggesting this approach.}. We estimated a cross-correlation coefficient $\rho^{\rm noise}_\ell$ of the noise in harmonic space from the ratio of the cross spectrum to the square roots of the auto spectra of NPIPE and 2018 SMICA maps, after subtracting the CMB signal spectrum obtained from the cross-spectrum of the NPIPE A and B splits. On small scales, we found an almost scale-independent $\rho^{\rm noise}_\ell \sim 0.8$, both in temperature and polarization, and both in the SMICA maps and the main CMB frequency channels. We then constructed paired SMICA 2018 noise simulations by adding a Gaussian component to our SMICA NPIPE noise simulation, so that the new 2018 simulation has the right auto-power and cross-correlation $\rho_\ell^{\rm noise}$ to the NPIPE one. Performing lensing reconstruction on the set of simulation built in this way, we can assess the significance (in this crude model) of the shifts compared to the NPIPE ones. The measured correlation value of 0.8 is smaller than expected from only 10\% more data, which suggests that this estimate may be conservative. We find that the lowest bin on the peak is $3.1 \sigma$ low according to this criterion for the MV reconstruction. For the TT reconstruction, the first two bins are about $2.4 \sigma$ low. Other points do not stand out.

\begin{table}
      \caption[]{bandpowers on the conservative range, for our Minimum Variance (MV) baseline, including joint temperature and polarization filtering. Our polarization-only (PP) reconstruction can be found in Table~\ref{table:tablePP}. The reconstruction uses inhomogeneous CMB filtering and the additional $\kappa$-filtering step. Numbers are given for $10^7 \cdot L^{2}(L + 1)^2 C_L^{\phi\phi} / 2\pi$.}
         \label{table:tableMV}
     $$
         \begin{array}{rcrll}
            \hline
            \noalign{\smallskip}
            L_{\rm min}& L_{\rm av}&L_{\rm max}      &  \textrm{MV-fid} & _{\rm fid}\hat A^{\phi^{\rm MV}} \\
            \noalign{\smallskip}
            \hline \hline
            \noalign{\smallskip}
  8 & 28.1 &  40 & 1.40 & 0.95 \pm 0.08 \\
 41 & 63.5 &  84 & 1.28 & 1.05 \pm 0.05 \\
 85 & 106.2 & 129 & 9.93 \cdot 10^{-1} & 1.01 \pm 0.04 \\
130 & 150.4 & 174 & 7.62 \cdot 10^{-1} & 0.91 \pm 0.05 \\
175 & 195.1 & 219 & 5.99 \cdot 10^{-1} & 0.92 \pm 0.06 \\
220 & 240.3 & 264 & 4.84 \cdot 10^{-1} & 0.94 \pm 0.08 \\
265 & 285.8 & 309 & 4.01 \cdot 10^{-1} & 1.10 \pm 0.10 \\
310 & 331.4 & 354 & 3.38 \cdot 10^{-1} & 1.10 \pm 0.12 \\
355 & 377.3 & 400 & 2.88 \cdot 10^{-1} & 0.86 \pm 0.14 \\
\noalign{\smallskip}
            \hline
         \end{array}
     $$
   \end{table}

  \begin{figure*}
   \centering
   \includegraphics[width=\textwidth]{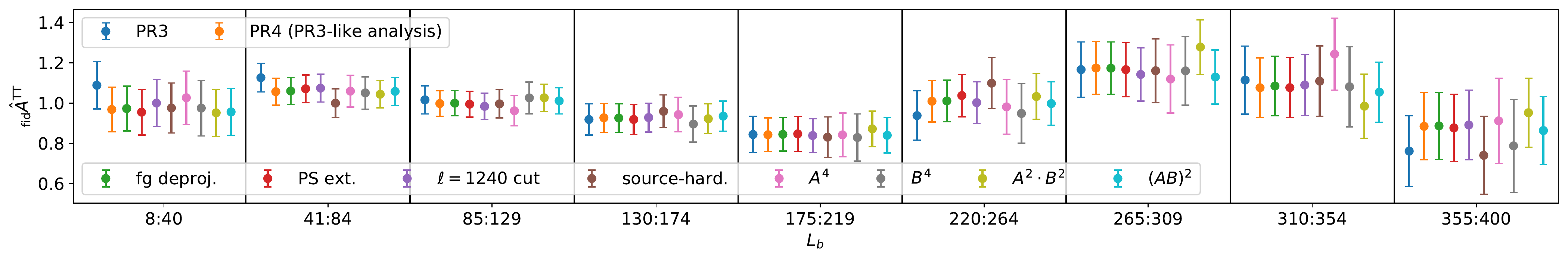}
   \caption{Lensing temperature-only bandpowers in units of the fiducial FFP10 lensing spectrum for each bin of the conservative range (without point-source subtraction), with the \PRthree~points in blue, and several test reconstructions on \PRfour~maps. All of the reconstructions used here use the same filtering and main analysis choices as \PRthree, with the orange points showing the \PRfour~result corresponding to the blue points. The filtering deprojects the residual foreground model to give the green points. The point-source component of the analysis mask is extended for the red points. The purple points exclude a range of 20 CMB multipoles centred on 1240, for reasons given in the text. The brown points show the reconstruction bias-hardened against point sources. The shifts seen in these cases are all compatible with the expected variance obtained from differencing simulations. Finally, the last four sets of points show results from the A and B splits, namely $\hat \kappa^{AA}\cdot \hat \kappa^{AA} $, $\hat \kappa^{BB}\cdot \hat \kappa^{BB} $, $\hat \kappa^{AA}\cdot \hat \kappa^{BB} $ and $\hat \kappa^{AB}\cdot \hat \kappa^{AB} $ respectively.\label{fig:TTvariations}}
    \end{figure*}

 Fig.~\ref{fig:TTvariations} shows the  \PRthree~(blue)~and \PRfour~(orange) conservative range reconstructions from temperature only, together with several \PRfour~variations, always using homogeneous filtering as for \PRthree. The points shown include
\begin{itemize}
	\item \textbf{fg deproj.}: The residual foreground model after SMICA weighting (see Fig.~\ref{fig:fgs} bottom panel) is assigned infinite variance in the filtering process to project it out. This is very efficient at reducing the (already small) common component to all simulations.
	\item \textbf{PS ext.}: The point-source component of the lensing analysis mask is extended, increasing the masked sky fraction by 1.5\%. This mask is based on resolved point sources at 143GHz and 217GHz, but a handful of bright sources can be seen leaking outside of it slightly after SMICA weighting both on the foreground model and data.
	\item \textbf{\boldsymbol{$\ell=1240 $} cut}: This excludes the CMB multipole range $1230 \leq \ell \leq 1250$ from the quadratic estimators. Motivations for this were the presence of a spike at $\ell \sim 1240$ in the otherwise smoother residual foreground model power spectrum seen on the lensing mask. This multipole also coincides with a break in the \PRthree~(hence by construction also our \PRfour) SMICA weights.
	\item \textbf{source-hard.}: This shows the lensing reconstruction bias-hardened against point-source contamination~\cite{Namikawa2013,Osborne:2013nna}. Here a correction is made at the map level, subtracting from the lensing estimate a properly normalized estimate of a point source signature, itself obtained with a quadratic estimator. This has been shown to be very effective against contaminants that act similarly to a local lensing magnification, at the cost of some signal to noise. In this case a couple of larger shifts can be seen in the bandpowers. However, the estimators themselves differ more than in the previous cases, and these shifts are no larger than the spread expected from simulations.
\end{itemize}
The last four sets of points in Fig.~\ref{fig:TTvariations} show the reconstructions from \PRfour~A and B data splits. We obtain lensing maps from the $A$ and $B$ splits and reconstruct the power from $\hat \kappa^{AA} \cdot\hat \kappa^{AA}$ (pink), $\hat \kappa^{BB} \cdot\hat \kappa^{BB}$ (grey), as well as $\hat \kappa^{AA} \cdot \hat \kappa^{BB}$ (chartreuse) and $\hat \kappa^{AB} \cdot \hat \kappa^{AB}$ (cyan). The last two have in principle vanishing contributions of the noise maps to their $N^{(0)}$ bias or mean-field power respectively.
We refer to \PL~for many more such consistency tests for \PRthree. We collect a couple of further points of comparison between the \PRthree~and \PRfour~results:
\begin{itemize}
\item $L=2$ anomaly: the \PL~reconstruction has a clearly anomalous-looking quadrupole. In the case of the MV reconstruction
\begin{equation}
	_{\rm{fid}}\hat A^{\phi\phi, \rm MV}_{L=2} = 22 \pm 3 \textrm{   (\PRthree)}
\end{equation}
This value, and the ones below, do not include a MC correction, which would not affect this discussion. We also neglect the non-Gaussianity of the quadrupole power posterior. This high quadrupole value is completely driven by temperature, with
\begin{equation}
_{\rm{fid}}\hat A^{\phi\phi, \rm TT}_{L=2} = 40 \pm 4\textrm{   (\PRthree)},
\end{equation}
The much noisier polarization-only reconstruction is consistent with zero. We note that the mean-field spectrum is maximal at very low multipoles and absolutely massive (in these units, the TT mean-field is ${\sim} 5\cdot 10^4$ at $L=2$, but only ${\sim} 16$ on the first conservative bin $8 \leq L \leq 40$), so that accurate reconstruction is challenging and can be ruined by small inaccuracies in the simulation modelling. On the other hand, we now see different values on the NPIPE maps. Using the same analysis as \PL\ we find
\begin{equation}
   	_{\rm{fid}}\hat A^{\phi\phi, \rm  MV}_{L=2} = 6 \pm 2.5,\textrm{   (\PRfour, 2018-like)},
\end{equation}
and
\begin{equation}
   	_{\rm{fid}}\hat A^{\phi\phi, \rm  TT}_{L=2} = 8 \pm 4,\textrm{   (\PRfour, 2018-like)},\\
\end{equation}
and with our new baseline analysis
\begin{equation}
		_{\rm{fid}}\hat A^{\phi\phi, \rm GMV}_{L=2} = 1.3 \pm 1.9, \textrm{  (\PRfour, $\kappa$-filtered)}
\end{equation}
This better-behaved temperature reconstruction result seems robust against choices of data splits and combinations, as shown in Fig.~\ref{fig:verylowLs}. In all cases we use the same filtering choices as \PL.
We have also performed reconstruction using the full mission maps, with the SMICA-weighted Commander foreground template either de-projected in the filtering step, or subtracted from the maps prior to filtering, with very consistent results.

\begin{figure}
   \centering
   \includegraphics[width=\hsize]{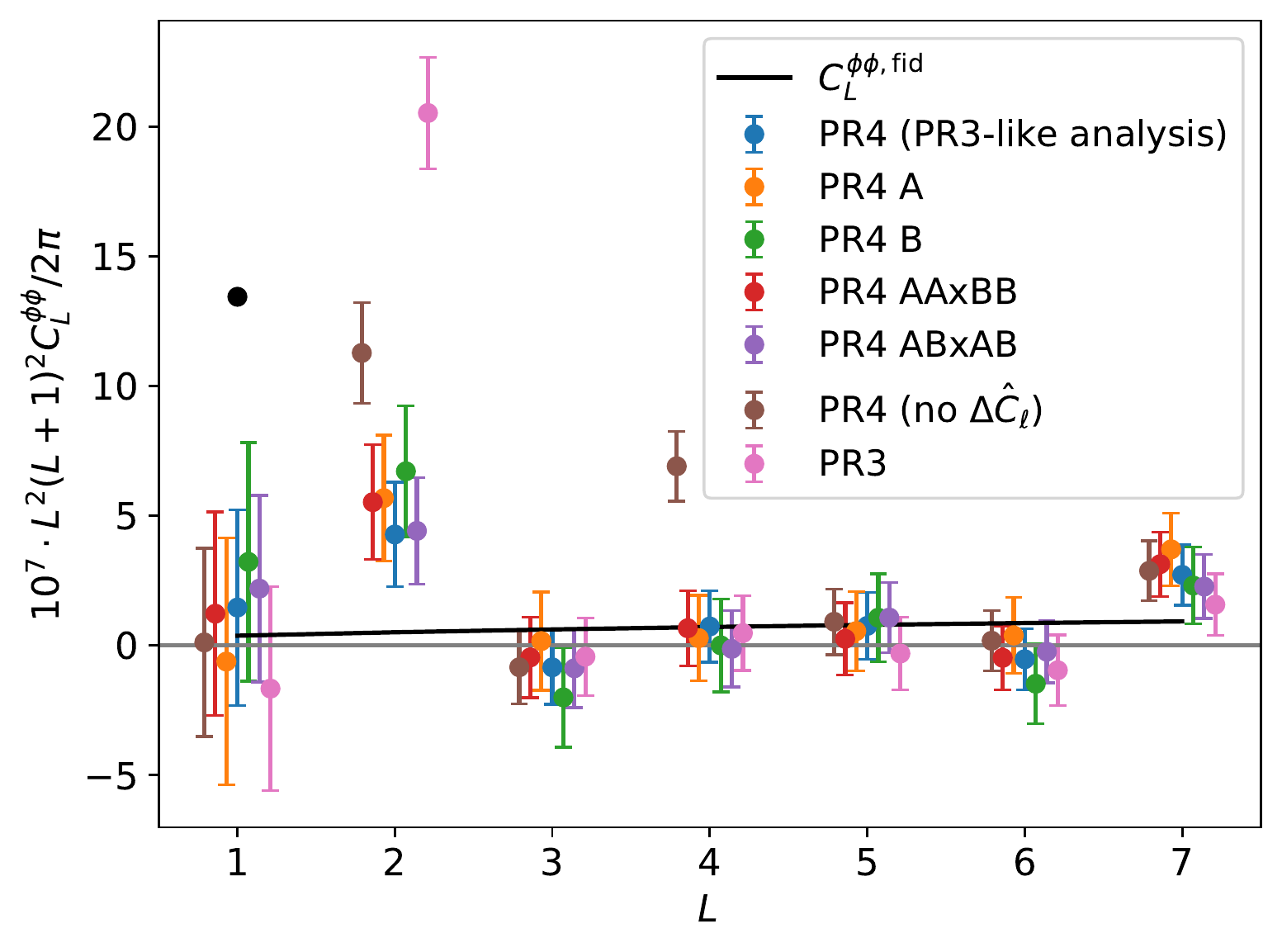}
      \caption{Lensing spectrum reconstruction from temperature on the lowest multipoles for a variety of NPIPE data splits and combinations. The comparison result for \PL~ (pink) clearly shows an anomalous-looking quadrupole (all points are shown here without Monte-Carlo corrections, which do not solve this issue). The brown points show the effect of not adding the few-percent sized additional high-$\ell$ power to the temperature simulations to match the unresolved foreground power in the data. If the missing power can be treated as originating from a Gaussian field, the signature on the bandpowers is quadratic in the mismatch, but the mean field at the first couple of even multipoles is so large that it becomes distinctly visible. The signature is however completely negligible on all of the conservative range bins. The black dot shows the expected dipolar aberration signal from our motion with respect to the CMB frame, which appears to be consistently taken into account by the mean-field in both cases. None of these very-low lensing multipoles enter the lensing likelihood.
         \label{fig:verylowLs}}
   \end{figure}

It is tempting to speculate from this that the NPIPE processing and simulations handle better at least some of the largest-scale non-idealities, though more concrete evidence is lacking to back this up further.
\begin{figure}
   \centering
   \includegraphics[width=\hsize]{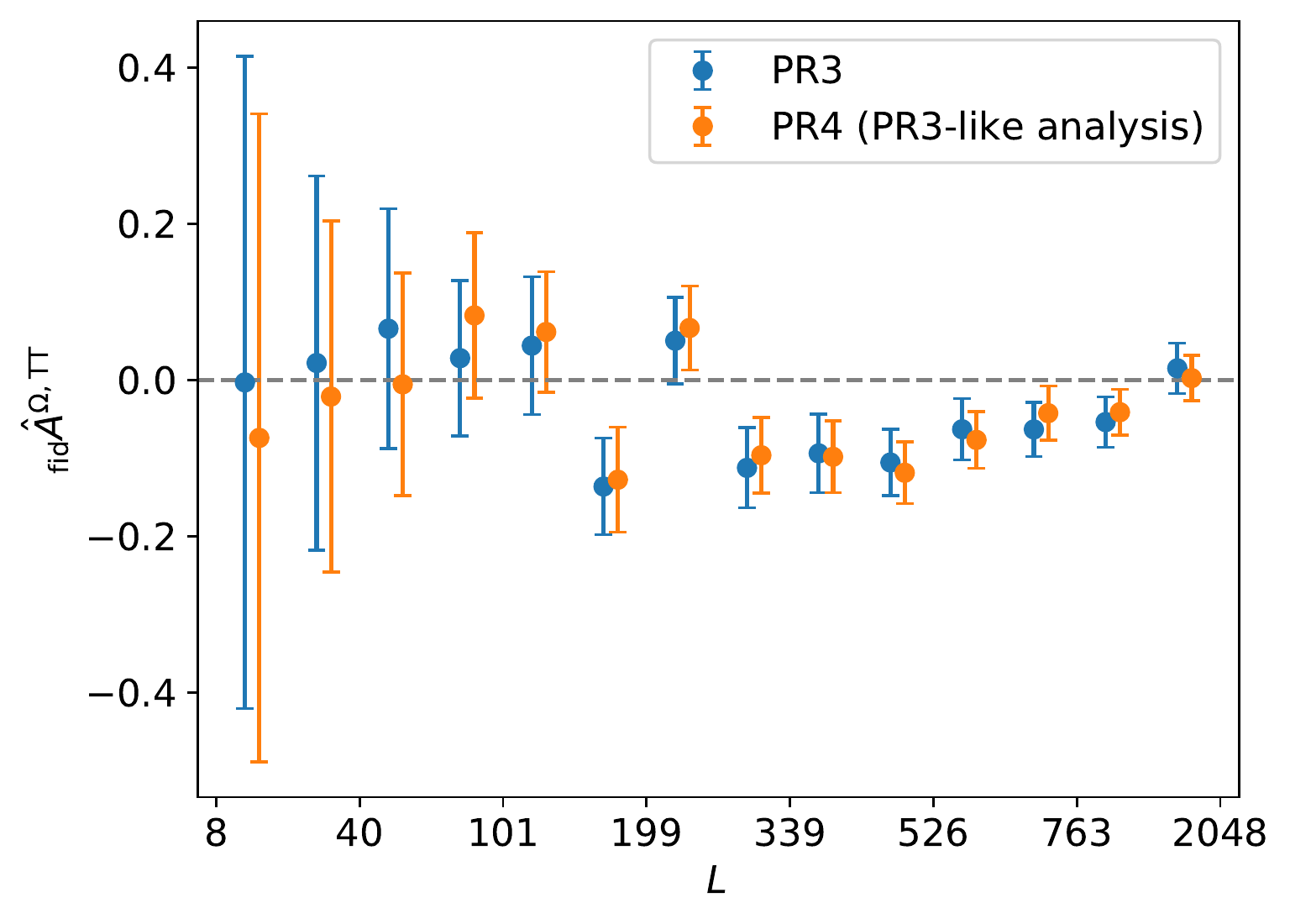}
      \caption{Lensing curl mode spectrum in units of a fiducial spectrum $L^2(L+1)^2C_L^{\Omega\Omega} /2\pi = 10^{-7}$, for the 2018 analysis and the same analysis on the new \PRfour~maps. The $4.3\sigma$ significance of the low feature at $264 \leq L \leq 901$ as assessed by \PL~remains virtually unchanged with the NPIPE new data processing. Reconstructions on the \PRfour~A and B splits give very similar results.\label{fig:xttcomp}}
   \end{figure}
\item Lensing curl spectrum from temperature: \PL~discussed and tested in some detail the presence of a possibly suspicious feature in the high-$L$ lensing curl mode constructed from temperature. In that regard, however, the NPIPE results are very similar. Fig.~\ref{fig:xttcomp} shows the $\rm{TT}$-curl mode spectrum, processed with the same homogeneous filtering pipeline. The reonstructions from the splits give qualitatively similar results.
\end{itemize}

\subsection{Polarization-only reconstruction}\label{sec:polonly}

   \begin{table}
      \caption[]{bandpowers on the conservative range, for our polarization-only reconstruction. Our Minimum variance bandpowers can be found in Table~\ref{table:tableMV}. This reconstruction uses inhomogeneous CMB filtering and the additional $\kappa$-filtering step. Numbers are given for $10^7\cdot L^{2}(L + 1)^2 C_L^{\phi\phi} / 2\pi$.}
         \label{table:tablePP}
     $$
         \begin{array}{rcrll}
            \hline
            \noalign{\smallskip}
            L_{\rm min}& L_{\rm av}&L_{\rm max}      &  \textrm{PP-fid} & _{\rm fid}\hat A^{\phi^{\rm PP}} \\
            \noalign{\smallskip}
            \hline \hline
            \noalign{\smallskip}
  8 & 28.0 &  40 & 1.40 & 0.67 \pm 0.23 \\
 41 & 63.3 &  84 & 1.28 & 0.94 \pm 0.14 \\
 85 & 105.8 & 129 & 9.95 \cdot 10^{-1} & 0.94 \pm 0.15 \\
130 & 149.8 & 174 & 7.64 \cdot 10^{-1} & 0.88 \pm 0.19 \\
175 & 194.3 & 219 & 6.01 \cdot 10^{-1} & 0.85 \pm 0.26 \\
220 & 239.3 & 264 & 4.86 \cdot 10^{-1} & 1.18 \pm 0.36 \\
265 & 285.0 & 309 & 4.02 \cdot 10^{-1} & 0.89 \pm 0.56 \\
310 & 330.9 & 354 & 3.38 \cdot 10^{-1} & 0.30 \pm 0.62 \\
355 & 377.0 & 400 & 2.89 \cdot 10^{-1} & 0.33 \pm 0.82 \\
\noalign{\smallskip}
            \hline
         \end{array}
     $$
   \end{table}

Table~\ref{table:tablePP} lists our updated polarization bandpowers. Compressing this into an amplitude, we find
\begin{equation}
	_{\rm{fid}}\hat A^{\phi\phi, \rm PP}_{8\rightarrow400} = 0.89 \pm 0.07\textrm{   (\PRfour, $\kappa$-filtered)},
\end{equation}
whereas the corresponding number quoted in \PL~is (with inhomogeneous CMB-filtering) $0.95 \pm 0.11$. The improvement in signal to noise can be decomposed roughly as follows. The lensing reconstruction noise is proportional to two powers of the data spectra. While the instrumental noise sources only 10\% of the lensing reconstruction noise for temperature, this dominates for polarization. Hence the reduced noise of the NPIPE maps has a greater impact on the polarization-only reconstruction. Using the \PRfour~maps in place of \PRthree\ we see a decrease in bandpower variance of almost 25\% (on the peak) to 35\% (on the tail). The benefit of $\kappa$-filtering is (as expected) more subtle, but still visible on the first two bins of the conservative range, with an improvement of ${\sim} 8\%$. For this polarization reconstruction, we have also pre-processed the maps, by modifying the SMICA weights in order to flatten the sub-degree scales empirical noise spectra. By making the fiducial covariance model of the CMB closer to that of the data, this also slightly decreases the error bars, by about $6\%$.
\subsection{ISW-lensing}
\begin{figure}
   \centering
   \includegraphics[width=\hsize]{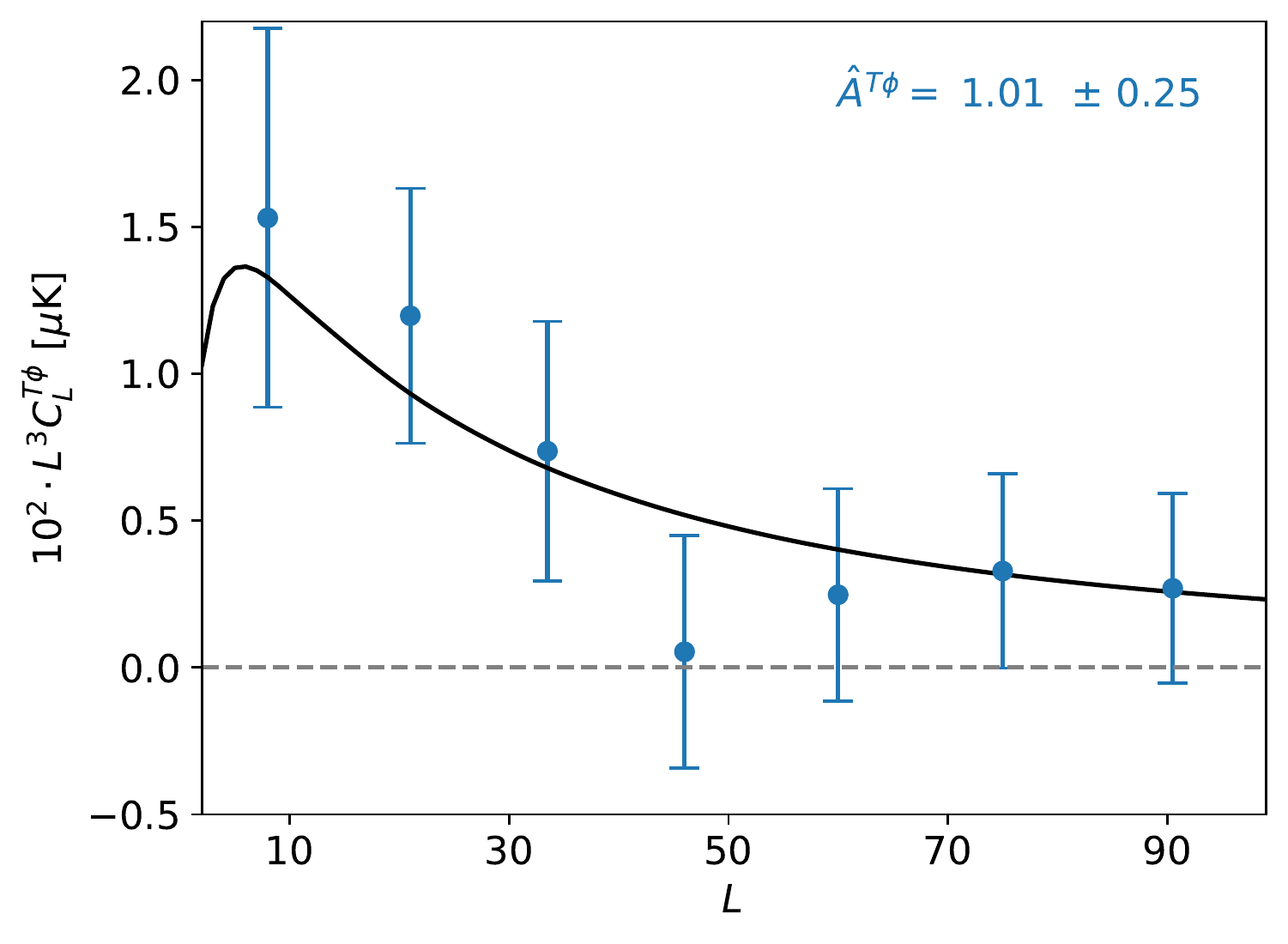}
      \caption{Estimate of the ISW lensing bispectrum $\hat C_L^{T \phi}$, on the multipole range $3 \leq L \leq 100$, from the cross-spectrum of the homogeneously-filtered MV NPIPE lensing reconstruction with the filtered temperature map. The black line shows the prediction in the fiducial cosmology. The detection significance is formally just above $4\sigma$.}
         \label{fig:ISWlensing}
  \end{figure}

Cross-correlating the lensing quadratic estimator to the temperature gives the lensing ISW bispectrum~\citep{Zaldarriaga:2000ud,Hu:2000ee,Lewis:2011fk}, which is visible on large scales. For simplicity we evaluate this only using the simplest homogenously-filtered lensing reconstruction map.
In Fig.~\ref{fig:ISWlensing} we show our estimate of $\hat C_\ell^{T \phi}$, which we obtained from the simple estimator
\begin{equation}
	\hat C_L^{T \phi} = \frac{1}{(2L + 1) f_{\rm sky}} \sum_{M = -L}^L\hat \phi^{\rm MV}_{LM} \hat T^*_{LM},
\end{equation}
where $\hat T$ is our homogeneously inverse-variance Wiener-filtered temperature map $T^{\rm WF}$, rescaled with the isotropic limit $C^{TT, \rm{fid}}_\ell / (C^{TT, \rm fid}_\ell + N^{TT, \rm fid}_\ell)$ of the filter in order to make it unbiased away from the mask. In principle, we could use the Wiener-filtered lensing maps here as well, but the improvement is negligible. We also apply an empirical Monte-Carlo correction to the bandpowers, calculated as the ratio of the mean $\hat C_L^{T\phi}$ across the simulation set to the FFP10 simulation input. This is at most a $4$-$5\%$ correction in a couple of bins, and generally a tiny fraction of the error bars. The approximate Gaussian covariance formula $(C_L^{\phi\phi, \rm fid} + N_L^{(0)}) C_L^{TT, \rm fid}/ (2L + 1) $ matches the empirical variance shape as a function of $L$, and we use it to inverse-weight $\hat C_L^{T \phi}$ and to build an estimate of the detection significance. We get
\begin{align}
	\: _{\rm fid}\hat A^{T\phi} = 1.01 \pm 0.25,
\end{align}
non-zero at just about 4$\sigma$. This is a bit higher than the value quoted in Ref.~\cite{Planck:2015mym} (just above $3\sigma$), from a combination of $10\%$ tighter error bars and a slightly higher recovered amplitude. The same analysis on the \PRthree~maps give $_{\rm fid}\hat A^{T\phi}=0.94 \pm 0.30$.
Finally, we note that $C_L^{E\phi}$ remains undetectable by a large margin and we see $\hat C_L^{E\phi}$ perfectly consistent with zero.

\subsection{Parameter constraints}\label{sec:parameters}
\begin{figure}
   \centering
   \includegraphics[width=\hsize]{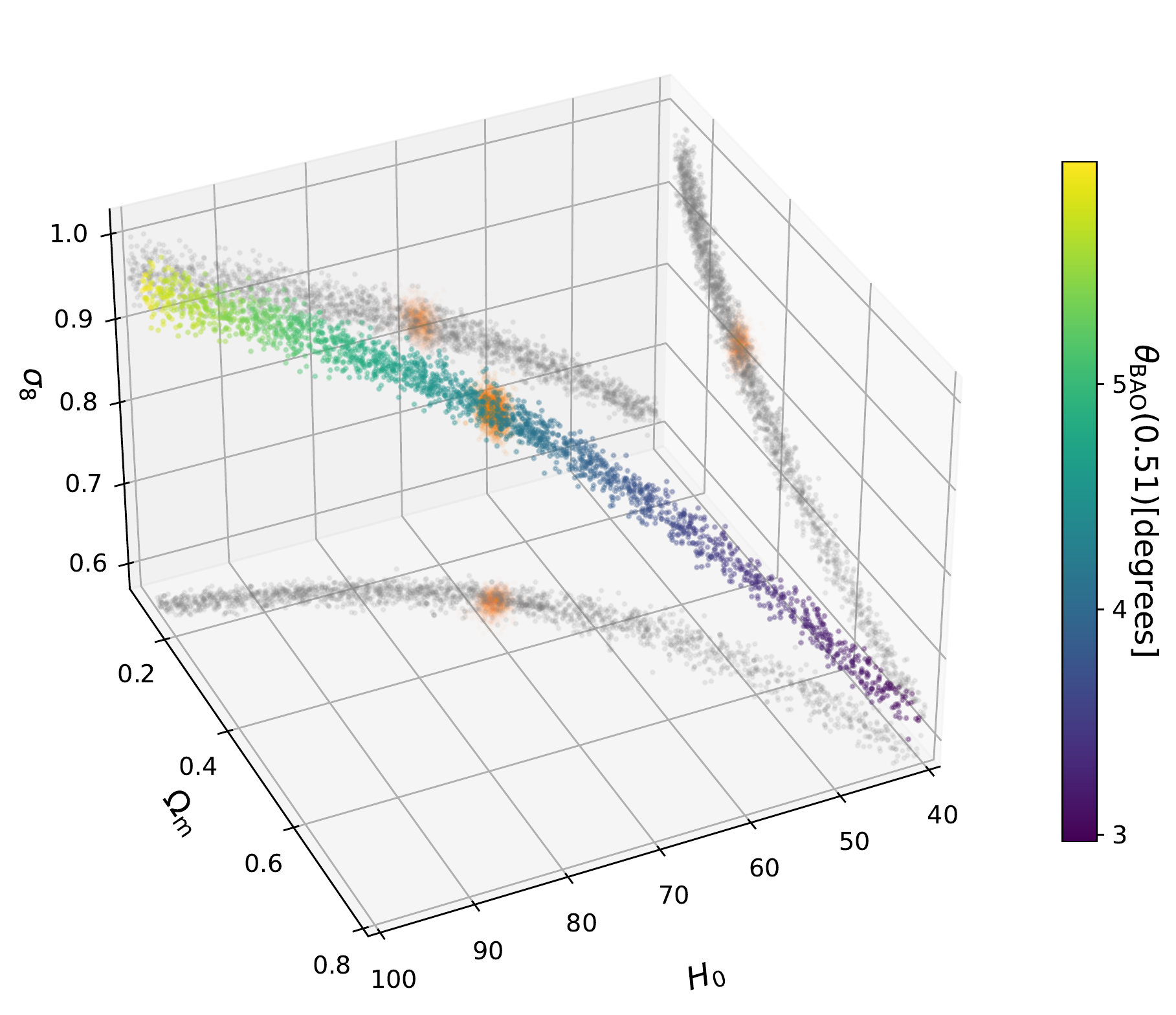}
   \includegraphics[width=\hsize]{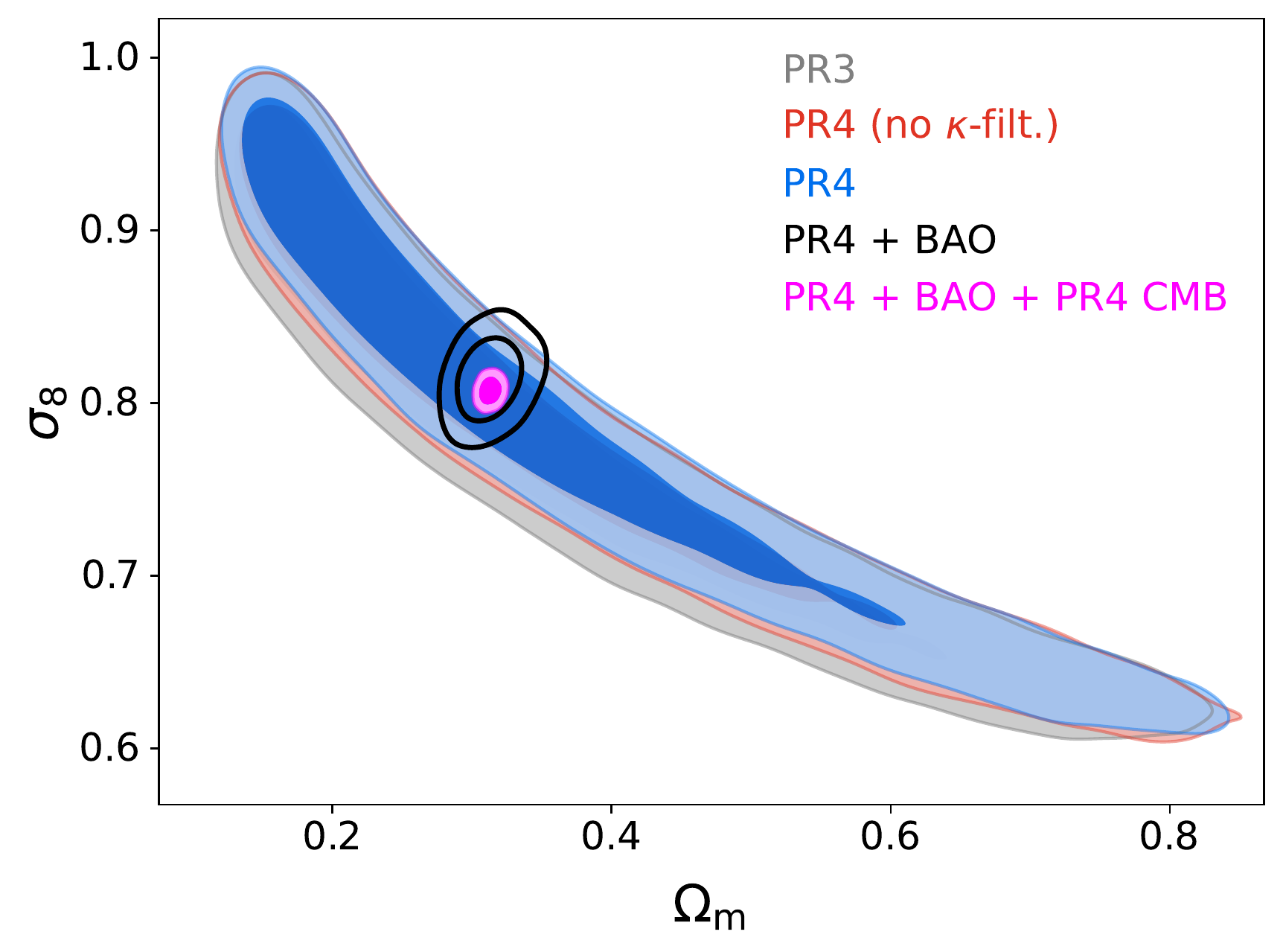}

      \caption{
      \textit{Upper panel:} Posterior samples in $H_0$-$\sigma_8$-$\Omega_{\mathrm{m}}$ space from our lensing-only likelihood, colour-coded by the angular size of the BAO scale $\theta_{\rm{BAO}}(z)\equiv r_{\rm{d}}/D_M(z)$ at redshift ${\sim} 0.5$, where $r_{\rm{d}}$ and $D_M$ are the comoving sound horizon and comoving distance respectively. The $H_0$ values are in km s$^{-1}$ Mpc$^{-1}$. The grey points show the projection into the 2D subspaces, and the orange points the posterior samples when combined with SDSS BAO data. \\
      \textit{Lower panel:} Constraints in the $\sigma_8$-$\Omega_{\mathrm{m}}$ plane from our lensing-only likelihood, with and without $\kappa$-filtering (blue and red respectively) and in comparison to \PL(grey). The black contours show the constraints combining our lensing likelihood with BAO data, and the magenta contours also include the recent \PRfour~based CMB spectra likelihood. Summary statistics are given in Tables~\ref{table:paramvars1} and~\ref{table:paramvars2}, and throughout the text.}
         \label{fig:s8omegam}
  \end{figure}

We use the Markov Chain Monte Carlo sampling method with Cobaya\footnote{\url{https://github.com/CobayaSampler/cobaya}}~\cite{Torrado:2020dgo} to estimate cosmological parameters from the calculated likelihoods, and analyse the resulting chains using GetDist\footnote{\url{https://github.com/cmbant/getdist}}~\cite{Lewis:2019xzd}. We use our baseline GMV, $\kappa$-filtered NPIPE CMB-marginalized likelihood (labelled PR4) with and without Baryon Accoustic Oscillations (BAO) likelihoods, and also the likelihood of the NPIPE analysis without applying the $\kappa$ filter with \PL priors. We generally follow the parameter definitions and priors of \PL, but use updated BAO likelihoods and an updated baryon density ($\Omega_{\rm b} h^2$) prior based on works that were published since \PL, as described in Tables~\ref{table:paramvars1} and~\ref{table:paramvars2}. The resulting base $\Lambda$CDM model parameter constraints, which we now discuss, are shown in these tables and Fig.~\ref{fig:s8omegam}.\\

	As shown in the upper panel of Fig.~\ref{fig:s8omegam}, in the base $\Lambda$CDM model the CMB lensing-only parameter degeneracies follow a narrow tube in the space spanned by $\sigma_8$-$H_0$-$\Omega_{\mathrm{m}}$ (see the detailed discussion of Ref.~\cite{Ade:2015zua} on the dependence of the lensing on cosmological parameters). This projects into a tight constraint on $\sigma_8\Omega^{0.25}$, as displayed in the lower panel. Our updated bandpowers induce a slight shift upward in the parameter mean compared to \PRthree, and a reduction in the 68\% confidence level, giving
	\begin{equation}
		\sigma_8 \Omega_{\rm m}^{0.25} = 0.599 \pm 0.016 \quad (\textrm{PR4}),
	\end{equation}
	where we had $0.589 \pm 0.020$ for \PRthree. The upward shift of the mean is consistent with the slightly \emph{lower} spectrum peak seen in \PRfour: As can be seen in Fig.~\ref{fig:sigmas}, for our noise levels the spectrum is best constrained just after the peak, so that the parameter fit is tightly constrained to the data there, in contrast to the other points. This, combined with the (more intuitively, positive) correlation between the amplitude of the tail and $\sigma_8 \Omega_{\rm m}^{0.25}$, produces an anti-correlation between the peak and tail. This anti-correlation is the same as that of the deflection amplitude (sourced disproportionately by the peak) to $\sigma_8 \Omega_{\rm m}^{0.25}$, studied in detail in Ref.~\cite{Ade:2015zua} [see Figs. 8 and 9 in particular]. In order to directly confirm this, we ran a chain with the first \PRfour~bin replaced by its \PRthree~value, which resulted in a 5 times smaller shift in $\sigma_8 \Omega_{\rm m}^{0.25}$, as expected.

       \begin{table}
   \renewcommand{\arraystretch}{1.8}
      \caption[]{Constraints on the CMB lensing parameter for different lensing datasets. The first row shows values from \PL for comparison.
      	We set $\Omega_{\rm{b}}h^2 = 0.02233 \pm 0.00036$ as a tight prior on the baryon density from recent element abundances and nucleosynthesis (BBN) modelling~\cite{Cooke:2017cwo, Mossa:2020gjc}, a Gaussian prior on $n_s=0.96\pm 0.02$, and the Hubble constant parameter is restricted to $0.4 < h < 1$.
      See Table~\ref{table:paramvars2} for results combined with BAO.}
         \label{table:paramvars1}
     $$
         \begin{array}{|l|c|}
			\hline
			\textrm{Lensing-only likelihood}	& \sigma_8 \Omega_{\mathrm{m}}^{0.25}  \\
			\hline \hline
				\textrm{PR3} 	& 0.589 \pm 0.020
								\\ \arrayrulecolor{gray!50}\hline\arrayrulecolor{black}
				\textrm{PR4} & 0.599 \pm 0.016  \\
				\textrm{PR4} \textrm{, no } \kappa\textrm{-filtering} & 0.596 \pm 0.017 \\
				\textrm{PR4, polarization-only} & 0.618 \pm 0.076 \\
			\hline
         \end{array}
     $$
   \end{table}
    \begin{table}
   \renewcommand{\arraystretch}{1.8}
      \caption[]{Constraints on the main lensing-related $\Lambda$CDM cosmological parameters for different lensing datasets, when combining BAO data with the lensing likelihood. The first row shows values from \PL for comparison.
      The BAO likelihoods for \PRfour~use the 6dF Galaxy Survey measurements~\cite{Beutler:2011hx}, the SDSS Data Release (DR) 7 measurements~\cite{Ross:2014qpa} and the angular and radial DR12+DR16 LRG data from eBOSS~\citep{eBOSS:2020yzd}, with the exception of the fourth entry, which uses the same BAO likelihoods and priors of \PRthree. We use the same priors described in Table~\ref{table:paramvars1} (except again for the fourth entry),
      and $H_0$ is in units of km s$^{-1}$ Mpc$^{-1}$.}
         \label{table:paramvars2}
$$
\hspace*{-0.15cm}         \begin{array}{|l|ccc|}
			\hline \textrm{Lensing}+\textrm{BAO}  & \sigma_8 & H_0 & \Omega_{\mathrm{m}}\\
			\hline \hline
				\textrm{PR3}& 0.811 \pm 0.019  & 67.9^{+1.2}_{-1.3} & 0.303^{+0.016}_{-0.018} \\ \arrayrulecolor{gray!50}\hline\arrayrulecolor{black}
				\textrm{PR4}  & 0.814 \pm 0.016 & 68.14^{+0.99}_{-1.10} & 0.313^{+0.014}_{-0.016} \\
				\textrm{PR4, no $\kappa$-filtering}& 0.810 \pm 0.016 & 68.2^{+1.0}_{-1.2} & 0.313^{+0.015}_{-0.017} \\
				\textrm{PR4} \textrm{ (PR3 priors \& BAO)} & 0.816 \pm 0.017 & 68.4^{+1.0}_{-1.3} & 0.309^{+0.014}_{-0.017} \\
				\textrm{PR4, polarization-only} & 0.814^{+0.051}_{-0.046} & 71.0^{+2.1}_{-2.9} & 0.363^{+0.036}_{-0.047} \\
			\hline
         \end{array}
     $$
   \end{table}

The points in the upper panel of Fig.~\ref{fig:s8omegam} are colour-coded by the BAO angular scale at redshift $z\sim 0.5$ (the BBN prior on the baryon density that we use means that the comoving sound horizon can be predicted accurately). A measurement of the projected sound horizon from BAO measurements at low redshift therefore breaks the three-way degeneracy, without using data from CMB spectra, giving tight constraints on the parameters individually. The constraints are shown in Table~\ref{table:paramvars2}. We find a similar decrease in errors bars compared to \PRthree\ as above, with very similar $H_0$ and $\sigma_8$ central values, and the entirety of the small data shift being seemingly transferred to a higher $\Omega_{\rm m}$ value. However, the fourth row of the table shows that at least part of this shift is due to the updated priors and BAO likelihoods. In the lensing + BAO chains of Table~\ref{table:paramvars2} the estimated galaxy weak-lensing parameter $S_8 \equiv \sigma_8 \left(\Omega_{\rm m}/0.3\right)^{0.5}$ central value is shifted upwards from $0.815 \pm 0.036$ (\PRthree) to $0.831\pm 0.029$ (\PRfour), with a very similar increase seen in $S_8$ from \PRthree~to \PRfour\ using \PRthree\ priors \& BAO.

Ref.~\cite{Rosenberg:2022sdy} recently extended the \planck~\texttt{CamSpec} high$-\ell$ CMB spectra likelihood to \PRfour, with overall consistency to \PRthree~and ${\sim} 10\%$ more constraining power compared to a comparable analysis with \PRthree. It also shifts the beyond $\Lambda$CDM parameters $A_L$ and $\Omega_K$ closer to their $\Lambda $CDM values (these two parameters have a tendency to peak slightly away from $\Lambda $CDM values without the inclusion of lensing data in \PRthree). Combining this new likelihood with lensing and BAO (while dropping the informative priors on $\Omega_{\rm b} h^2$ and $n_{\rm s}$, and adding the PR3 low-$\ell$ TT and EE likelihoods), gives the magenta confidence contours in Fig.~\ref{fig:s8omegam}. Both $\sigma_8$ and $H_0$ are then constrained well below the percent level,
\begin{equation}\left.\begin{matrix}
	\sigma_8 &=& 0.8072 \pm 0.0054 \\
	 H_0 &=& 67.41 \pm 0.39 \\
	 \Omega_{\rm m} &=& 0.3130 \pm 0.0053
\end{matrix}\right\} \textrm{PR4 + BAO + PR4 CMB,}
\end{equation}
where $H_0$ is in km s$^{-1}$ Mpc$^{-1}$. The central values are slightly shifted but consistent with the 2018 \planck~release results, with errors that are ${\sim} 15{-}25\%$ tighter (largely driven by the 10\% improvement from the PR4 high-$\ell$ CMB~\cite{Rosenberg:2022sdy} and the additional sky area used by the \texttt{CamSpec} likelihood compared to the original 2018 release~\cite{Efstathiou:2019mdh}).

We also considered extending $\Lambda$CDM to include a varying total neutrino mass, but the slightly improved lensing spectrum is not able to make much of a difference to constraints from the CMB spectra + BAO alone. We find
\begin{equation}
\underbrace{\sum m_\nu < 0.133 \textrm{ eV}  \textrm{  (95\% CL upper bound)}}_{\textrm{PR4 + BAO + PR4 CMB}}.
\end{equation}
With BAO and lensing only (PR4 + BAO), the constraint is obviously much weaker, finding $\sum m_\nu < 2.23 \textrm{ eV, (95\% CL upper bound)}$.
\\ Finally, tables~\ref{table:paramvars1} and \ref{table:paramvars2} also give the constraints using our lensing likelihood built from polarization only, showing consistency with of course larger errors. These errors are factors of 2 to 3 smaller than those achievable  on this same data without inhomogeneous filtering and the improvements discussed in this work.

\section{Conclusions}
We have presented an updated CMB lensing reconstruction from \planck~data, performed on the \PRfour~(NPIPE) maps. In part due to the lower noise in the CMB maps, and in part due to a small series of lensing pipeline improvements, error bars on the bandpowers have decreased, and constraint on the main lensing parameter has tightened to $\sigma_8\Omega_{\mathrm{m}}^{0.25} = 0.599 \pm 0.016$ from $0.589 \pm 0.020$, with an upward shift in mean value caused by a slightly lower lensing spectrum peak amplitude. This upward shift in lensing parameter is not unexpected given the change in error bars~\cite{Gratton:2019fru}, and its anti-correlation with the peak height for \planck~noise levels. The shift in bandpowers at the lensing spectrum peak is more substantial (1$\sigma$). From the available simulations of \PRthree~and \PRfour, it seems difficult to assess precisely the true significance of this peak shift, but we argued it is plausibly unexpected at the $3\sigma$ level given only statistical errors. At the lowest lensing multipoles, \PRfour~now gives robust results where \PRthree~had an anomalous-looking quadrupole.
	
	Apart from the $1\sigma$ peak shift, our lensing results are consistent with \PL. In particular, the broad region of low lensing curl power at $264 \leq L \leq 901$ that was a potential source of worry in \PRthree, with a quoted significance of $2.9\sigma$ ($4.2\sigma$ without consideration of look-elsewhere effects), is still very much there. As in \PL, for our likelihood analysis we restricted the bandpowers to a conservative multipole range that excludes much of this range.
	
	We have enhanced the \PRthree~lensing reconstruction and likelihood pipelines to account for the inhomogeneity of the lensing reconstruction noise across the sky, and improved our joint temperature and polarization filtering scheme. These novel more-optimal techniques account for about half of our improvement in error bars. Apart from the polarization-only reconstruction, which is greatly improved but still provides weak constraints (a 7\% constraint on the lensing power amplitude) compared to temperature, the impact of the analysis improvements is relatively modest for \planck~data.  It could be substantially higher for future experiments, particularly for ones with inhomogeneous scan strategies.	

Our new lensing likelihoods with and without CMB-marginalization are available in a Cobaya-friendly format at \url{https://github.com/carronj/planck_PR4_lensing}. Maps and simulations are available at NERSC as indicated at this same page. Other lensing map variations can be made available upon request to the first author.

\begin{acknowledgements}
JC acknowledges support from a SNSF Eccellenza Professorial Fellowship (No. 186879),
and AL support by the UK STFC grant ST/T000473/1.
We thank Reijo Keskitalo for exchanges and help regarding the \PRfour~maps. We thank Kareem Marzouk and Sebastian Belkner for related collaboration and work on the \PRfour~maps and SMICA pipeline~\cite{Marzouk:2022utf}. We thank Erik Rosenberg for sharing the PR4 \texttt{CamSpec} likelihood~\cite{Rosenberg:2022sdy}. We thank Anthony Challinor and more generally the entire lensing working group of Simons Observatory for discussions at various stages of this work. JC thanks Louis Legrand for useful comments on the draft. This research used resources of the National Energy Research Scientific Computing Center (NERSC), a U.S. Department of Energy Office of Science User Facility located at Lawrence Berkeley National Laboratory. Some of the results in this paper have been derived using the healpy and HEALPix\footnote{\url{http://healpix.sourceforge.net}}package~\cite{Zonca2019, Gorski:2004by}.
\end{acknowledgements}

\appendix
\bibliography{julbib,texbase/antony,texbase/cosmomc}

\providecommand{\aj}{Astron. J. }\providecommand{\apj}{ApJ
  }\providecommand{\apjl}{ApJ
  }\providecommand{\mnras}{MNRAS}\providecommand{\prl}{PRL}\providecommand{\prd}{PRD}\providecommand{\jcap}{JCAP}\providecommand{\aap}{A\&A}
\begin{thebibliography}{51}%
\makeatletter
\providecommand \@ifxundefined [1]{%
 \@ifx{#1\undefined}
}%
\providecommand \@ifnum [1]{%
 \ifnum #1\expandafter \@firstoftwo
 \else \expandafter \@secondoftwo
 \fi
}%
\providecommand \@ifx [1]{%
 \ifx #1\expandafter \@firstoftwo
 \else \expandafter \@secondoftwo
 \fi
}%
\providecommand \natexlab [1]{#1}%
\providecommand \enquote  [1]{``#1''}%
\providecommand \bibnamefont  [1]{#1}%
\providecommand \bibfnamefont [1]{#1}%
\providecommand \citenamefont [1]{#1}%
\providecommand \href@noop [0]{\@secondoftwo}%
\providecommand \href [0]{\begingroup \@sanitize@url \@href}%
\providecommand \@href[1]{\@@startlink{#1}\@@href}%
\providecommand \@@href[1]{\endgroup#1\@@endlink}%
\providecommand \@sanitize@url [0]{\catcode `\\12\catcode `\$12\catcode
  `\&12\catcode `\#12\catcode `\^12\catcode `\_12\catcode `\%12\relax}%
\providecommand \@@startlink[1]{}%
\providecommand \@@endlink[0]{}%
\providecommand \url  [0]{\begingroup\@sanitize@url \@url }%
\providecommand \@url [1]{\endgroup\@href {#1}{\urlprefix }}%
\providecommand \urlprefix  [0]{URL }%
\providecommand \Eprint [0]{\href }%
\providecommand \doibase [0]{https://doi.org/}%
\providecommand \selectlanguage [0]{\@gobble}%
\providecommand \bibinfo  [0]{\@secondoftwo}%
\providecommand \bibfield  [0]{\@secondoftwo}%
\providecommand \translation [1]{[#1]}%
\providecommand \BibitemOpen [0]{}%
\providecommand \bibitemStop [0]{}%
\providecommand \bibitemNoStop [0]{.\EOS\space}%
\providecommand \EOS [0]{\spacefactor3000\relax}%
\providecommand \BibitemShut  [1]{\csname bibitem#1\endcsname}%
\let\auto@bib@innerbib\@empty
\bibitem [{\citenamefont {Aghanim}\ \emph
  {et~al.}(2020{\natexlab{a}})\citenamefont {Aghanim} \emph
  {et~al.}}]{Aghanim:2018oex}%
  \BibitemOpen
  \bibfield  {author} {\bibinfo {author} {\bibfnamefont {N.}~\bibnamefont
  {Aghanim}} \emph {et~al.} (\bibinfo {collaboration} {Planck}),\ }\bibfield
  {title} {\bibinfo {title} {{Planck 2018 results. VIII. Gravitational
  lensing}},\ }\href {https://doi.org/10.1051/0004-6361/201833886} {\bibfield
  {journal} {\bibinfo  {journal} {Astron. Astrophys.}\ }\textbf {\bibinfo
  {volume} {641}},\ \bibinfo {pages} {A8} (\bibinfo {year}
  {2020}{\natexlab{a}})},\ \Eprint {https://arxiv.org/abs/1807.06210}
  {arXiv:1807.06210 [astro-ph.CO]} \BibitemShut {NoStop}%
\bibitem [{\citenamefont {Sherwin}\ \emph {et~al.}(2016)\citenamefont {Sherwin}
  \emph {et~al.}}]{Sherwin:2016tyf}%
  \BibitemOpen
  \bibfield  {author} {\bibinfo {author} {\bibfnamefont {B.~D.}\ \bibnamefont
  {Sherwin}} \emph {et~al.} (\bibinfo {collaboration} {ACT}),\ }\bibfield
  {title} {\bibinfo {title} {{The Atacama Cosmology Telescope: Two-Season
  ACTPol Lensing Power Spectrum}},\ }\href@noop {} {\bibfield  {journal}
  {\bibinfo  {journal} {Submitted to: Phys. Rev. D}\ } (\bibinfo {year}
  {2016})},\ \Eprint {https://arxiv.org/abs/1611.09753} {arXiv:1611.09753
  [astro-ph.CO]} \BibitemShut {NoStop}%
\bibitem [{\citenamefont {Bianchini}\ \emph {et~al.}(2020)\citenamefont
  {Bianchini} \emph {et~al.}}]{SPT:2019fqo}%
  \BibitemOpen
  \bibfield  {author} {\bibinfo {author} {\bibfnamefont {F.}~\bibnamefont
  {Bianchini}} \emph {et~al.} (\bibinfo {collaboration} {SPT}),\ }\bibfield
  {title} {\bibinfo {title} {{Constraints on Cosmological Parameters from the
  500 deg$^2$ SPTpol Lensing Power Spectrum}},\ }\href
  {https://doi.org/10.3847/1538-4357/ab6082} {\bibfield  {journal} {\bibinfo
  {journal} {Astrophys. J.}\ }\textbf {\bibinfo {volume} {888}},\ \bibinfo
  {pages} {119} (\bibinfo {year} {2020})},\ \Eprint
  {https://arxiv.org/abs/1910.07157} {arXiv:1910.07157 [astro-ph.CO]}
  \BibitemShut {NoStop}%
\bibitem [{\citenamefont {Wu}\ \emph {et~al.}(2019)\citenamefont {Wu} \emph
  {et~al.}}]{Wu:2019hek}%
  \BibitemOpen
  \bibfield  {author} {\bibinfo {author} {\bibfnamefont {W.~L.~K.}\
  \bibnamefont {Wu}} \emph {et~al.},\ }\bibfield  {title} {\bibinfo {title} {{A
  Measurement of the Cosmic Microwave Background Lensing Potential and Power
  Spectrum from 500 deg$^2$ of SPTpol Temperature and Polarization Data}},\
  }\href {https://doi.org/10.3847/1538-4357/ab4186} {\bibfield  {journal}
  {\bibinfo  {journal} {Astrophys. J.}\ }\textbf {\bibinfo {volume} {884}},\
  \bibinfo {pages} {70} (\bibinfo {year} {2019})},\ \Eprint
  {https://arxiv.org/abs/1905.05777} {arXiv:1905.05777 [astro-ph.CO]}
  \BibitemShut {NoStop}%
\bibitem [{\citenamefont {Darwish}\ \emph {et~al.}(2020)\citenamefont {Darwish}
  \emph {et~al.}}]{Darwish:2020fwf}%
  \BibitemOpen
  \bibfield  {author} {\bibinfo {author} {\bibfnamefont {O.}~\bibnamefont
  {Darwish}} \emph {et~al.},\ }\bibfield  {title} {\bibinfo {title} {{The
  Atacama Cosmology Telescope: A CMB lensing mass map over 2100 square degrees
  of sky and its cross-correlation with BOSS-CMASS galaxies}},\ }\href
  {https://doi.org/10.1093/mnras/staa3438} {\bibfield  {journal} {\bibinfo
  {journal} {Mon. Not. Roy. Astron. Soc.}\ }\textbf {\bibinfo {volume} {500}},\
  \bibinfo {pages} {2250} (\bibinfo {year} {2020})},\ \Eprint
  {https://arxiv.org/abs/2004.01139} {arXiv:2004.01139 [astro-ph.CO]}
  \BibitemShut {NoStop}%
\bibitem [{\citenamefont {Akrami}\ \emph
  {et~al.}(2020{\natexlab{a}})\citenamefont {Akrami} \emph
  {et~al.}}]{Planck:2020olo}%
  \BibitemOpen
  \bibfield  {author} {\bibinfo {author} {\bibfnamefont {Y.}~\bibnamefont
  {Akrami}} \emph {et~al.} (\bibinfo {collaboration} {Planck}),\ }\bibfield
  {title} {\bibinfo {title} {{$Planck$ intermediate results. LVII. Joint Planck
  LFI and HFI data processing}},\ }\href
  {https://doi.org/10.1051/0004-6361/202038073} {\bibfield  {journal} {\bibinfo
   {journal} {Astron. Astrophys.}\ }\textbf {\bibinfo {volume} {643}},\
  \bibinfo {pages} {A42} (\bibinfo {year} {2020}{\natexlab{a}})},\ \Eprint
  {https://arxiv.org/abs/2007.04997} {arXiv:2007.04997 [astro-ph.CO]}
  \BibitemShut {NoStop}%
\bibitem [{\citenamefont {Okamoto}\ and\ \citenamefont
  {Hu}(2003)}]{Okamoto:2003zw}%
  \BibitemOpen
  \bibfield  {author} {\bibinfo {author} {\bibfnamefont {T.}~\bibnamefont
  {Okamoto}}\ and\ \bibinfo {author} {\bibfnamefont {W.}~\bibnamefont {Hu}},\
  }\bibfield  {title} {\bibinfo {title} {{CMB lensing reconstruction on the
  full sky}},\ }\href {https://doi.org/10.1103/PhysRevD.67.083002} {\bibfield
  {journal} {\bibinfo  {journal} {Phys. Rev.}\ }\textbf {\bibinfo {volume}
  {D67}},\ \bibinfo {pages} {083002} (\bibinfo {year} {2003})},\ \Eprint
  {https://arxiv.org/abs/astro-ph/0301031} {arXiv:astro-ph/0301031 [astro-ph]}
  \BibitemShut {NoStop}%
\bibitem [{\citenamefont {Hu}\ and\ \citenamefont {Okamoto}(2002)}]{Hu:2001kj}%
  \BibitemOpen
  \bibfield  {author} {\bibinfo {author} {\bibfnamefont {W.}~\bibnamefont
  {Hu}}\ and\ \bibinfo {author} {\bibfnamefont {T.}~\bibnamefont {Okamoto}},\
  }\bibfield  {title} {\bibinfo {title} {{Mass reconstruction with {CMB}
  polarization}},\ }\href {https://doi.org/10.1086/341110} {\bibfield
  {journal} {\bibinfo  {journal} {\apj}\ }\textbf {\bibinfo {volume} {574}},\
  \bibinfo {pages} {566} (\bibinfo {year} {2002})},\ \Eprint
  {https://arxiv.org/abs/astro-ph/0111606} {arXiv:astro-ph/0111606 [astro-ph]}
  \BibitemShut {NoStop}%
\bibitem [{\citenamefont {Maniyar}\ \emph {et~al.}(2021)\citenamefont
  {Maniyar}, \citenamefont {Ali-Ha\"\i{}moud}, \citenamefont {Carron},
  \citenamefont {Lewis},\ and\ \citenamefont
  {Madhavacheril}}]{Maniyar:2021msb}%
  \BibitemOpen
  \bibfield  {author} {\bibinfo {author} {\bibfnamefont {A.~S.}\ \bibnamefont
  {Maniyar}}, \bibinfo {author} {\bibfnamefont {Y.}~\bibnamefont
  {Ali-Ha\"\i{}moud}}, \bibinfo {author} {\bibfnamefont {J.}~\bibnamefont
  {Carron}}, \bibinfo {author} {\bibfnamefont {A.}~\bibnamefont {Lewis}},\ and\
  \bibinfo {author} {\bibfnamefont {M.~S.}\ \bibnamefont {Madhavacheril}},\
  }\bibfield  {title} {\bibinfo {title} {{Quadratic estimators for CMB weak
  lensing}},\ }\href {https://doi.org/10.1103/PhysRevD.103.083524} {\bibfield
  {journal} {\bibinfo  {journal} {Phys. Rev. D}\ }\textbf {\bibinfo {volume}
  {103}},\ \bibinfo {pages} {083524} (\bibinfo {year} {2021})},\ \Eprint
  {https://arxiv.org/abs/2101.12193} {arXiv:2101.12193 [astro-ph.CO]}
  \BibitemShut {NoStop}%
\bibitem [{\citenamefont {Aguirre}\ \emph {et~al.}(2019)\citenamefont {Aguirre}
  \emph {et~al.}}]{Ade:2018sbj}%
  \BibitemOpen
  \bibfield  {author} {\bibinfo {author} {\bibfnamefont {J.}~\bibnamefont
  {Aguirre}} \emph {et~al.} (\bibinfo {collaboration} {Simons Observatory}),\
  }\bibfield  {title} {\bibinfo {title} {{The Simons Observatory: Science goals
  and forecasts}},\ }\href {https://doi.org/10.1088/1475-7516/2019/02/056}
  {\bibfield  {journal} {\bibinfo  {journal} {JCAP}\ }\textbf {\bibinfo
  {volume} {1902}},\ \bibinfo {pages} {056}},\ \Eprint
  {https://arxiv.org/abs/1808.07445} {arXiv:1808.07445 [astro-ph.CO]}
  \BibitemShut {NoStop}%
\bibitem [{\citenamefont {Mirmelstein}\ \emph {et~al.}(2019)\citenamefont
  {Mirmelstein}, \citenamefont {Carron},\ and\ \citenamefont
  {Lewis}}]{Mirmelstein:2019sxi}%
  \BibitemOpen
  \bibfield  {author} {\bibinfo {author} {\bibfnamefont {M.}~\bibnamefont
  {Mirmelstein}}, \bibinfo {author} {\bibfnamefont {J.}~\bibnamefont
  {Carron}},\ and\ \bibinfo {author} {\bibfnamefont {A.}~\bibnamefont
  {Lewis}},\ }\bibfield  {title} {\bibinfo {title} {{Optimal filtering for CMB
  lensing reconstruction}},\ }\href
  {https://doi.org/10.1103/PhysRevD.100.123509} {\bibfield  {journal} {\bibinfo
   {journal} {Phys. Rev. D}\ }\textbf {\bibinfo {volume} {100}},\ \bibinfo
  {pages} {123509} (\bibinfo {year} {2019})},\ \Eprint
  {https://arxiv.org/abs/1909.02653} {arXiv:1909.02653 [astro-ph.CO]}
  \BibitemShut {NoStop}%
\bibitem [{\citenamefont {Adam}\ \emph {et~al.}(2016)\citenamefont {Adam} \emph
  {et~al.}}]{Planck:2015mis}%
  \BibitemOpen
  \bibfield  {author} {\bibinfo {author} {\bibfnamefont {R.}~\bibnamefont
  {Adam}} \emph {et~al.} (\bibinfo {collaboration} {Planck}),\ }\bibfield
  {title} {\bibinfo {title} {{Planck 2015 results. IX. Diffuse component
  separation: CMB maps}},\ }\href {https://doi.org/10.1051/0004-6361/201525936}
  {\bibfield  {journal} {\bibinfo  {journal} {Astron. Astrophys.}\ }\textbf
  {\bibinfo {volume} {594}},\ \bibinfo {pages} {A9} (\bibinfo {year} {2016})},\
  \Eprint {https://arxiv.org/abs/1502.05956} {arXiv:1502.05956 [astro-ph.CO]}
  \BibitemShut {NoStop}%
\bibitem [{\citenamefont {Akrami}\ \emph
  {et~al.}(2020{\natexlab{b}})\citenamefont {Akrami} \emph
  {et~al.}}]{Planck:2018yye}%
  \BibitemOpen
  \bibfield  {author} {\bibinfo {author} {\bibfnamefont {Y.}~\bibnamefont
  {Akrami}} \emph {et~al.} (\bibinfo {collaboration} {Planck}),\ }\bibfield
  {title} {\bibinfo {title} {{Planck 2018 results. IV. Diffuse component
  separation}},\ }\href {https://doi.org/10.1051/0004-6361/201833881}
  {\bibfield  {journal} {\bibinfo  {journal} {Astron. Astrophys.}\ }\textbf
  {\bibinfo {volume} {641}},\ \bibinfo {pages} {A4} (\bibinfo {year}
  {2020}{\natexlab{b}})},\ \Eprint {https://arxiv.org/abs/1807.06208}
  {arXiv:1807.06208 [astro-ph.CO]} \BibitemShut {NoStop}%
\bibitem [{\citenamefont {Cardoso}\ \emph {et~al.}(2008)\citenamefont
  {Cardoso}, \citenamefont {Martin}, \citenamefont {Delabrouille},
  \citenamefont {Betoule},\ and\ \citenamefont {Patanchon}}]{Cardoso:2008qt}%
  \BibitemOpen
  \bibfield  {author} {\bibinfo {author} {\bibfnamefont {J.-F.}\ \bibnamefont
  {Cardoso}}, \bibinfo {author} {\bibfnamefont {M.}~\bibnamefont {Martin}},
  \bibinfo {author} {\bibfnamefont {J.}~\bibnamefont {Delabrouille}}, \bibinfo
  {author} {\bibfnamefont {M.}~\bibnamefont {Betoule}},\ and\ \bibinfo {author}
  {\bibfnamefont {G.}~\bibnamefont {Patanchon}},\ }\bibfield  {title} {\bibinfo
  {title} {{Component separation with flexible models. Application to the
  separation of astrophysical emissions}},\ }\href@noop {} {\  (\bibinfo {year}
  {2008})},\ \Eprint {https://arxiv.org/abs/0803.1814} {arXiv:0803.1814
  [astro-ph]} \BibitemShut {NoStop}%
\bibitem [{\citenamefont {Fernandez-Cobos}\ \emph {et~al.}(2012)\citenamefont
  {Fernandez-Cobos}, \citenamefont {Vielva}, \citenamefont {Barreiro},\ and\
  \citenamefont {Martinez-Gonzalez}}]{Fernandez-Cobos:2011mmt}%
  \BibitemOpen
  \bibfield  {author} {\bibinfo {author} {\bibfnamefont {R.}~\bibnamefont
  {Fernandez-Cobos}}, \bibinfo {author} {\bibfnamefont {P.}~\bibnamefont
  {Vielva}}, \bibinfo {author} {\bibfnamefont {R.~B.}\ \bibnamefont
  {Barreiro}},\ and\ \bibinfo {author} {\bibfnamefont {E.}~\bibnamefont
  {Martinez-Gonzalez}},\ }\bibfield  {title} {\bibinfo {title}
  {{Multi-resolution internal template cleaning: An application to the
  Wilkinson Microwave Anisotropy Probe 7-yr polarization data}},\ }\href
  {https://doi.org/10.1111/j.1365-2966.2011.20182.x} {\bibfield  {journal}
  {\bibinfo  {journal} {Mon. Not. Roy. Astron. Soc.}\ }\textbf {\bibinfo
  {volume} {420}},\ \bibinfo {pages} {2162} (\bibinfo {year} {2012})},\ \Eprint
  {https://arxiv.org/abs/1106.2016} {arXiv:1106.2016 [astro-ph.CO]}
  \BibitemShut {NoStop}%
\bibitem [{\citenamefont {Ade}\ \emph {et~al.}(2014{\natexlab{a}})\citenamefont
  {Ade} \emph {et~al.}}]{Planck:2013maj}%
  \BibitemOpen
  \bibfield  {author} {\bibinfo {author} {\bibfnamefont {P.~A.~R.}\
  \bibnamefont {Ade}} \emph {et~al.} (\bibinfo {collaboration} {Planck}),\
  }\bibfield  {title} {\bibinfo {title} {{Planck 2013 results. XIV. Zodiacal
  emission}},\ }\href {https://doi.org/10.1051/0004-6361/201321562} {\bibfield
  {journal} {\bibinfo  {journal} {Astron. Astrophys.}\ }\textbf {\bibinfo
  {volume} {571}},\ \bibinfo {pages} {A14} (\bibinfo {year}
  {2014}{\natexlab{a}})},\ \Eprint {https://arxiv.org/abs/1303.5074}
  {arXiv:1303.5074 [astro-ph.CO]} \BibitemShut {NoStop}%
\bibitem [{\citenamefont {Ade}\ \emph {et~al.}(2016{\natexlab{a}})\citenamefont
  {Ade} \emph {et~al.}}]{Planck:2015mym}%
  \BibitemOpen
  \bibfield  {author} {\bibinfo {author} {\bibfnamefont {P.~A.~R.}\
  \bibnamefont {Ade}} \emph {et~al.} (\bibinfo {collaboration} {Planck}),\
  }\bibfield  {title} {\bibinfo {title} {{Planck 2015 results. XV.
  Gravitational lensing}},\ }\href
  {https://doi.org/10.1051/0004-6361/201525941} {\bibfield  {journal} {\bibinfo
   {journal} {Astron. Astrophys.}\ }\textbf {\bibinfo {volume} {594}},\
  \bibinfo {pages} {A15} (\bibinfo {year} {2016}{\natexlab{a}})},\ \Eprint
  {https://arxiv.org/abs/1502.01591} {arXiv:1502.01591 [astro-ph.CO]}
  \BibitemShut {NoStop}%
\bibitem [{\citenamefont {Hivon}\ \emph {et~al.}(2017)\citenamefont {Hivon},
  \citenamefont {Mottet},\ and\ \citenamefont {Ponthieu}}]{Hivon:2016qyw}%
  \BibitemOpen
  \bibfield  {author} {\bibinfo {author} {\bibfnamefont {E.}~\bibnamefont
  {Hivon}}, \bibinfo {author} {\bibfnamefont {S.}~\bibnamefont {Mottet}},\ and\
  \bibinfo {author} {\bibfnamefont {N.}~\bibnamefont {Ponthieu}},\ }\bibfield
  {title} {\bibinfo {title} {{QuickPol: Fast calculation of effective beam
  matrices for CMB polarization}},\ }\href
  {https://doi.org/10.1051/0004-6361/201629626} {\bibfield  {journal} {\bibinfo
   {journal} {Astron. Astrophys.}\ }\textbf {\bibinfo {volume} {598}},\
  \bibinfo {pages} {A25} (\bibinfo {year} {2017})},\ \Eprint
  {https://arxiv.org/abs/1608.08833} {arXiv:1608.08833 [astro-ph.CO]}
  \BibitemShut {NoStop}%
\bibitem [{\citenamefont {Aghanim}\ \emph
  {et~al.}(2020{\natexlab{b}})\citenamefont {Aghanim} \emph
  {et~al.}}]{Planck:2019nip}%
  \BibitemOpen
  \bibfield  {author} {\bibinfo {author} {\bibfnamefont {N.}~\bibnamefont
  {Aghanim}} \emph {et~al.} (\bibinfo {collaboration} {Planck}),\ }\bibfield
  {title} {\bibinfo {title} {{Planck 2018 results. V. CMB power spectra and
  likelihoods}},\ }\href {https://doi.org/10.1051/0004-6361/201936386}
  {\bibfield  {journal} {\bibinfo  {journal} {Astron. Astrophys.}\ }\textbf
  {\bibinfo {volume} {641}},\ \bibinfo {pages} {A5} (\bibinfo {year}
  {2020}{\natexlab{b}})},\ \Eprint {https://arxiv.org/abs/1907.12875}
  {arXiv:1907.12875 [astro-ph.CO]} \BibitemShut {NoStop}%
\bibitem [{\citenamefont {Smith}\ \emph {et~al.}(2007)\citenamefont {Smith},
  \citenamefont {Zahn},\ and\ \citenamefont {Dore}}]{Smith:2007rg}%
  \BibitemOpen
  \bibfield  {author} {\bibinfo {author} {\bibfnamefont {K.~M.}\ \bibnamefont
  {Smith}}, \bibinfo {author} {\bibfnamefont {O.}~\bibnamefont {Zahn}},\ and\
  \bibinfo {author} {\bibfnamefont {O.}~\bibnamefont {Dore}},\ }\bibfield
  {title} {\bibinfo {title} {{Detection of Gravitational Lensing in the Cosmic
  Microwave Background}},\ }\href {https://doi.org/10.1103/PhysRevD.76.043510}
  {\bibfield  {journal} {\bibinfo  {journal} {Phys. Rev.}\ }\textbf {\bibinfo
  {volume} {D76}},\ \bibinfo {pages} {043510} (\bibinfo {year} {2007})},\
  \Eprint {https://arxiv.org/abs/0705.3980} {arXiv:0705.3980 [astro-ph]}
  \BibitemShut {NoStop}%
\bibitem [{\citenamefont {Hanson}\ \emph {et~al.}(2011)\citenamefont {Hanson},
  \citenamefont {Challinor}, \citenamefont {Efstathiou},\ and\ \citenamefont
  {Bielewicz}}]{Hanson:2010rp}%
  \BibitemOpen
  \bibfield  {author} {\bibinfo {author} {\bibfnamefont {D.}~\bibnamefont
  {Hanson}}, \bibinfo {author} {\bibfnamefont {A.}~\bibnamefont {Challinor}},
  \bibinfo {author} {\bibfnamefont {G.}~\bibnamefont {Efstathiou}},\ and\
  \bibinfo {author} {\bibfnamefont {P.}~\bibnamefont {Bielewicz}},\ }\bibfield
  {title} {\bibinfo {title} {{CMB temperature lensing power reconstruction}},\
  }\href {https://doi.org/10.1103/PhysRevD.83.043005} {\bibfield  {journal}
  {\bibinfo  {journal} {\prd}\ }\textbf {\bibinfo {volume} {83}},\ \bibinfo
  {pages} {043005} (\bibinfo {year} {2011})},\ \Eprint
  {https://arxiv.org/abs/1008.4403} {arXiv:1008.4403 [astro-ph.CO]}
  \BibitemShut {NoStop}%
\bibitem [{\citenamefont {Kesden}\ \emph {et~al.}(2003)\citenamefont {Kesden},
  \citenamefont {Cooray},\ and\ \citenamefont {Kamionkowski}}]{Kesden:2003cc}%
  \BibitemOpen
  \bibfield  {author} {\bibinfo {author} {\bibfnamefont {M.}~\bibnamefont
  {Kesden}}, \bibinfo {author} {\bibfnamefont {A.}~\bibnamefont {Cooray}},\
  and\ \bibinfo {author} {\bibfnamefont {M.}~\bibnamefont {Kamionkowski}},\
  }\bibfield  {title} {\bibinfo {title} {Lensing reconstruction with {CMB}
  temperature and polarization},\ }\href@noop {} {\bibfield  {journal}
  {\bibinfo  {journal} {\prd}\ }\textbf {\bibinfo {volume} {67}},\ \bibinfo
  {pages} {123507} (\bibinfo {year} {2003})},\ \Eprint
  {https://arxiv.org/abs/astro-ph/0302536} {astro-ph/0302536} \BibitemShut
  {NoStop}%
\bibitem [{\citenamefont {Story}\ \emph {et~al.}(2015)\citenamefont {Story}
  \emph {et~al.}}]{Story:2014hni}%
  \BibitemOpen
  \bibfield  {author} {\bibinfo {author} {\bibfnamefont {K.~T.}\ \bibnamefont
  {Story}} \emph {et~al.} (\bibinfo {collaboration} {SPT}),\ }\bibfield
  {title} {\bibinfo {title} {{A Measurement of the Cosmic Microwave Background
  Gravitational Lensing Potential from 100 Square Degrees of SPTpol Data}},\
  }\href {https://doi.org/10.1088/0004-637X/810/1/50} {\bibfield  {journal}
  {\bibinfo  {journal} {Astrophys. J.}\ }\textbf {\bibinfo {volume} {810}},\
  \bibinfo {pages} {50} (\bibinfo {year} {2015})},\ \Eprint
  {https://arxiv.org/abs/1412.4760} {arXiv:1412.4760 [astro-ph.CO]}
  \BibitemShut {NoStop}%
\bibitem [{\citenamefont {B{\"o}hm}\ \emph {et~al.}(2016)\citenamefont
  {B{\"o}hm}, \citenamefont {Schmittfull},\ and\ \citenamefont
  {Sherwin}}]{Bohm:2016gzt}%
  \BibitemOpen
  \bibfield  {author} {\bibinfo {author} {\bibfnamefont {V.}~\bibnamefont
  {B{\"o}hm}}, \bibinfo {author} {\bibfnamefont {M.}~\bibnamefont
  {Schmittfull}},\ and\ \bibinfo {author} {\bibfnamefont {B.~D.}\ \bibnamefont
  {Sherwin}},\ }\bibfield  {title} {\bibinfo {title} {{Bias to CMB lensing
  measurements from the bispectrum of large-scale structure}},\ }\href
  {https://doi.org/10.1103/PhysRevD.94.043519} {\bibfield  {journal} {\bibinfo
  {journal} {Phys. Rev.}\ }\textbf {\bibinfo {volume} {D94}},\ \bibinfo {pages}
  {043519} (\bibinfo {year} {2016})},\ \Eprint
  {https://arxiv.org/abs/1605.01392} {arXiv:1605.01392 [astro-ph.CO]}
  \BibitemShut {NoStop}%
\bibitem [{\citenamefont {Böhm}\ \emph {et~al.}(2018)\citenamefont {Böhm},
  \citenamefont {Sherwin}, \citenamefont {Liu}, \citenamefont {Hill},
  \citenamefont {Schmittfull},\ and\ \citenamefont {Namikawa}}]{Bohm:2018omn}%
  \BibitemOpen
  \bibfield  {author} {\bibinfo {author} {\bibfnamefont {V.}~\bibnamefont
  {Böhm}}, \bibinfo {author} {\bibfnamefont {B.~D.}\ \bibnamefont {Sherwin}},
  \bibinfo {author} {\bibfnamefont {J.}~\bibnamefont {Liu}}, \bibinfo {author}
  {\bibfnamefont {J.~C.}\ \bibnamefont {Hill}}, \bibinfo {author}
  {\bibfnamefont {M.}~\bibnamefont {Schmittfull}},\ and\ \bibinfo {author}
  {\bibfnamefont {T.}~\bibnamefont {Namikawa}},\ }\bibfield  {title} {\bibinfo
  {title} {{Effect of non-Gaussian lensing deflections on CMB lensing
  measurements}},\ }\href {https://doi.org/10.1103/PhysRevD.98.123510}
  {\bibfield  {journal} {\bibinfo  {journal} {Phys. Rev.}\ }\textbf {\bibinfo
  {volume} {D98}},\ \bibinfo {pages} {123510} (\bibinfo {year} {2018})},\
  \Eprint {https://arxiv.org/abs/1806.01157} {arXiv:1806.01157 [astro-ph.CO]}
  \BibitemShut {NoStop}%
\bibitem [{\citenamefont {Beck}\ \emph {et~al.}(2018)\citenamefont {Beck},
  \citenamefont {Fabbian},\ and\ \citenamefont {Errard}}]{Beck:2018wud}%
  \BibitemOpen
  \bibfield  {author} {\bibinfo {author} {\bibfnamefont {D.}~\bibnamefont
  {Beck}}, \bibinfo {author} {\bibfnamefont {G.}~\bibnamefont {Fabbian}},\ and\
  \bibinfo {author} {\bibfnamefont {J.}~\bibnamefont {Errard}},\ }\bibfield
  {title} {\bibinfo {title} {{Lensing Reconstruction in Post-Born Cosmic
  Microwave Background Weak Lensing}},\ }\href
  {https://doi.org/10.1103/PhysRevD.98.043512} {\bibfield  {journal} {\bibinfo
  {journal} {Phys. Rev.}\ }\textbf {\bibinfo {volume} {D98}},\ \bibinfo {pages}
  {043512} (\bibinfo {year} {2018})},\ \Eprint
  {https://arxiv.org/abs/1806.01216} {arXiv:1806.01216 [astro-ph.CO]}
  \BibitemShut {NoStop}%
\bibitem [{\citenamefont {Fabbian}\ \emph {et~al.}(2019)\citenamefont
  {Fabbian}, \citenamefont {Lewis},\ and\ \citenamefont
  {Beck}}]{Fabbian:2019tik}%
  \BibitemOpen
  \bibfield  {author} {\bibinfo {author} {\bibfnamefont {G.}~\bibnamefont
  {Fabbian}}, \bibinfo {author} {\bibfnamefont {A.}~\bibnamefont {Lewis}},\
  and\ \bibinfo {author} {\bibfnamefont {D.}~\bibnamefont {Beck}},\ }\bibfield
  {title} {\bibinfo {title} {{CMB lensing reconstruction biases in
  cross-correlation with large-scale structure probes}},\ }\href
  {https://doi.org/10.1088/1475-7516/2019/10/057} {\bibfield  {journal}
  {\bibinfo  {journal} {JCAP}\ }\textbf {\bibinfo {volume} {1910}}\bibfield
  {number} {\bibinfo  {number} { (10)},\ \bibinfo {pages} {057}},\ }\Eprint
  {https://arxiv.org/abs/1906.08760} {arXiv:1906.08760 [astro-ph.CO]}
  \BibitemShut {NoStop}%
\bibitem [{\citenamefont {Pratten}\ and\ \citenamefont
  {Lewis}(2016)}]{Pratten:2016dsm}%
  \BibitemOpen
  \bibfield  {author} {\bibinfo {author} {\bibfnamefont {G.}~\bibnamefont
  {Pratten}}\ and\ \bibinfo {author} {\bibfnamefont {A.}~\bibnamefont
  {Lewis}},\ }\bibfield  {title} {\bibinfo {title} {{Impact of post-Born
  lensing on the CMB}},\ }\href {https://doi.org/10.1088/1475-7516/2016/08/047}
  {\bibfield  {journal} {\bibinfo  {journal} {JCAP}\ }\textbf {\bibinfo
  {volume} {1608}}\bibfield  {number} {\bibinfo  {number} { (08)},\ \bibinfo
  {pages} {047}},\ }\Eprint {https://arxiv.org/abs/1605.05662}
  {arXiv:1605.05662 [astro-ph.CO]} \BibitemShut {NoStop}%
\bibitem [{\citenamefont {Stevens}\ \emph {et~al.}(2018)\citenamefont {Stevens}
  \emph {et~al.}}]{Stevens:2018biw}%
  \BibitemOpen
  \bibfield  {author} {\bibinfo {author} {\bibfnamefont {J.~R.}\ \bibnamefont
  {Stevens}} \emph {et~al.},\ }\bibfield  {title} {\bibinfo {title} {{Designs
  for next generation CMB survey strategies from Chile}},\ }\href
  {https://doi.org/10.1117/12.2313898} {\bibfield  {journal} {\bibinfo
  {journal} {Proc. SPIE Int. Soc. Opt. Eng.}\ }\textbf {\bibinfo {volume}
  {10708}},\ \bibinfo {pages} {1070841} (\bibinfo {year} {2018})},\ \Eprint
  {https://arxiv.org/abs/1808.05131} {arXiv:1808.05131 [astro-ph.IM]}
  \BibitemShut {NoStop}%
\bibitem [{\citenamefont {Peloton}\ \emph {et~al.}(2017)\citenamefont
  {Peloton}, \citenamefont {Schmittfull}, \citenamefont {Lewis}, \citenamefont
  {Carron},\ and\ \citenamefont {Zahn}}]{Peloton:2016kbw}%
  \BibitemOpen
  \bibfield  {author} {\bibinfo {author} {\bibfnamefont {J.}~\bibnamefont
  {Peloton}}, \bibinfo {author} {\bibfnamefont {M.}~\bibnamefont
  {Schmittfull}}, \bibinfo {author} {\bibfnamefont {A.}~\bibnamefont {Lewis}},
  \bibinfo {author} {\bibfnamefont {J.}~\bibnamefont {Carron}},\ and\ \bibinfo
  {author} {\bibfnamefont {O.}~\bibnamefont {Zahn}},\ }\bibfield  {title}
  {\bibinfo {title} {{Full covariance of CMB and lensing reconstruction power
  spectra}},\ }\href {https://doi.org/10.1103/PhysRevD.95.043508} {\bibfield
  {journal} {\bibinfo  {journal} {Phys. Rev.}\ }\textbf {\bibinfo {volume}
  {D95}},\ \bibinfo {pages} {043508} (\bibinfo {year} {2017})},\ \Eprint
  {https://arxiv.org/abs/1611.01446} {arXiv:1611.01446 [astro-ph.CO]}
  \BibitemShut {NoStop}%
\bibitem [{\citenamefont {Osborne}\ \emph {et~al.}(2014)\citenamefont
  {Osborne}, \citenamefont {Hanson},\ and\ \citenamefont
  {Doré}}]{Osborne:2013nna}%
  \BibitemOpen
  \bibfield  {author} {\bibinfo {author} {\bibfnamefont {S.~J.}\ \bibnamefont
  {Osborne}}, \bibinfo {author} {\bibfnamefont {D.}~\bibnamefont {Hanson}},\
  and\ \bibinfo {author} {\bibfnamefont {O.}~\bibnamefont {Doré}},\ }\bibfield
   {title} {\bibinfo {title} {{Extragalactic Foreground Contamination in
  Temperature-based CMB Lens Reconstruction}},\ }\href
  {https://doi.org/10.1088/1475-7516/2014/03/024} {\bibfield  {journal}
  {\bibinfo  {journal} {\jcap}\ }\textbf {\bibinfo {volume} {1403}},\ \bibinfo
  {pages} {024} (\bibinfo {year} {2014})},\ \Eprint
  {https://arxiv.org/abs/1310.7547} {arXiv:1310.7547 [astro-ph.CO]}
  \BibitemShut {NoStop}%
\bibitem [{\citenamefont {Ade}\ \emph {et~al.}(2014{\natexlab{b}})\citenamefont
  {Ade} \emph {et~al.}}]{Ade:2013tyw}%
  \BibitemOpen
  \bibfield  {author} {\bibinfo {author} {\bibfnamefont {P.~A.~R.}\
  \bibnamefont {Ade}} \emph {et~al.} (\bibinfo {collaboration} {Planck}),\
  }\bibfield  {title} {\bibinfo {title} {{Planck 2013 results. XVII.
  Gravitational lensing by large-scale structure}},\ }\href
  {https://doi.org/10.1051/0004-6361/201321543} {\bibfield  {journal} {\bibinfo
   {journal} {Astron. Astrophys.}\ }\textbf {\bibinfo {volume} {571}},\
  \bibinfo {pages} {A17} (\bibinfo {year} {2014}{\natexlab{b}})},\ \Eprint
  {https://arxiv.org/abs/1303.5077} {arXiv:1303.5077 [astro-ph.CO]}
  \BibitemShut {NoStop}%
\bibitem [{\citenamefont {Aghanim}\ \emph
  {et~al.}(2020{\natexlab{c}})\citenamefont {Aghanim} \emph
  {et~al.}}]{Planck:2018vyg}%
  \BibitemOpen
  \bibfield  {author} {\bibinfo {author} {\bibfnamefont {N.}~\bibnamefont
  {Aghanim}} \emph {et~al.} (\bibinfo {collaboration} {Planck}),\ }\bibfield
  {title} {\bibinfo {title} {{Planck 2018 results. VI. Cosmological
  parameters}},\ }\href {https://doi.org/10.1051/0004-6361/201833910}
  {\bibfield  {journal} {\bibinfo  {journal} {Astron. Astrophys.}\ }\textbf
  {\bibinfo {volume} {641}},\ \bibinfo {pages} {A6} (\bibinfo {year}
  {2020}{\natexlab{c}})},\ \bibinfo {note} {[Erratum: Astron.Astrophys. 652, C4
  (2021)]},\ \Eprint {https://arxiv.org/abs/1807.06209} {arXiv:1807.06209
  [astro-ph.CO]} \BibitemShut {NoStop}%
\bibitem [{\citenamefont {{Namikawa}}\ \emph {et~al.}(2013)\citenamefont
  {{Namikawa}}, \citenamefont {{Hanson}},\ and\ \citenamefont
  {{Takahashi}}}]{Namikawa2013}%
  \BibitemOpen
  \bibfield  {author} {\bibinfo {author} {\bibfnamefont {T.}~\bibnamefont
  {{Namikawa}}}, \bibinfo {author} {\bibfnamefont {D.}~\bibnamefont
  {{Hanson}}},\ and\ \bibinfo {author} {\bibfnamefont {R.}~\bibnamefont
  {{Takahashi}}},\ }\bibfield  {title} {\bibinfo {title} {{Bias-hardened CMB
  lensing}},\ }\href {https://doi.org/10.1093/mnras/stt195} {\bibfield
  {journal} {\bibinfo  {journal} {\mnras}\ }\textbf {\bibinfo {volume} {431}},\
  \bibinfo {pages} {609} (\bibinfo {year} {2013})},\ \Eprint
  {https://arxiv.org/abs/1209.0091} {arXiv:1209.0091 [astro-ph.CO]}
  \BibitemShut {NoStop}%
\bibitem [{\citenamefont {Zaldarriaga}(2000)}]{Zaldarriaga:2000ud}%
  \BibitemOpen
  \bibfield  {author} {\bibinfo {author} {\bibfnamefont {M.}~\bibnamefont
  {Zaldarriaga}},\ }\bibfield  {title} {\bibinfo {title} {{Lensing of the CMB:
  Non-Gaussian aspects}},\ }\href {https://doi.org/10.1103/PhysRevD.62.063510}
  {\bibfield  {journal} {\bibinfo  {journal} {\prd}\ }\textbf {\bibinfo
  {volume} {62}},\ \bibinfo {pages} {063510} (\bibinfo {year} {2000})},\
  \Eprint {https://arxiv.org/abs/astro-ph/9910498} {arXiv:astro-ph/9910498}
  \BibitemShut {NoStop}%
\bibitem [{\citenamefont {Hu}(2000)}]{Hu:2000ee}%
  \BibitemOpen
  \bibfield  {author} {\bibinfo {author} {\bibfnamefont {W.}~\bibnamefont
  {Hu}},\ }\bibfield  {title} {\bibinfo {title} {{Weak lensing of the CMB: A
  harmonic approach}},\ }\href {https://doi.org/10.1103/PhysRevD.62.043007}
  {\bibfield  {journal} {\bibinfo  {journal} {\prd}\ }\textbf {\bibinfo
  {volume} {62}},\ \bibinfo {pages} {043007} (\bibinfo {year} {2000})},\
  \Eprint {https://arxiv.org/abs/astro-ph/0001303} {arXiv:astro-ph/0001303
  [astro-ph]} \BibitemShut {NoStop}%
\bibitem [{\citenamefont {Lewis}\ \emph {et~al.}(2011)\citenamefont {Lewis},
  \citenamefont {Challinor},\ and\ \citenamefont {Hanson}}]{Lewis:2011fk}%
  \BibitemOpen
  \bibfield  {author} {\bibinfo {author} {\bibfnamefont {A.}~\bibnamefont
  {Lewis}}, \bibinfo {author} {\bibfnamefont {A.}~\bibnamefont {Challinor}},\
  and\ \bibinfo {author} {\bibfnamefont {D.}~\bibnamefont {Hanson}},\
  }\bibfield  {title} {\bibinfo {title} {{The shape of the CMB lensing
  bispectrum}},\ }\href {https://doi.org/10.1088/1475-7516/2011/03/018}
  {\bibfield  {journal} {\bibinfo  {journal} {JCAP}\ }\textbf {\bibinfo
  {volume} {03}},\ \bibinfo {pages} {018}},\ \Eprint
  {https://arxiv.org/abs/1101.2234} {arXiv:1101.2234 [astro-ph.CO]}
  \BibitemShut {NoStop}%
\bibitem [{\citenamefont {Torrado}\ and\ \citenamefont
  {Lewis}(2021)}]{Torrado:2020dgo}%
  \BibitemOpen
  \bibfield  {author} {\bibinfo {author} {\bibfnamefont {J.}~\bibnamefont
  {Torrado}}\ and\ \bibinfo {author} {\bibfnamefont {A.}~\bibnamefont
  {Lewis}},\ }\bibfield  {title} {\bibinfo {title} {{Cobaya: Code for Bayesian
  Analysis of hierarchical physical models}},\ }\href
  {https://doi.org/10.1088/1475-7516/2021/05/057} {\bibfield  {journal}
  {\bibinfo  {journal} {\jcap}\ }\textbf {\bibinfo {volume} {05}},\ \bibinfo
  {pages} {057} (\bibinfo {year} {2021})},\ \Eprint
  {https://arxiv.org/abs/2005.05290} {arXiv:2005.05290 [astro-ph.IM]}
  \BibitemShut {NoStop}%
\bibitem [{\citenamefont {Lewis}(2019)}]{Lewis:2019xzd}%
  \BibitemOpen
  \bibfield  {author} {\bibinfo {author} {\bibfnamefont {A.}~\bibnamefont
  {Lewis}},\ }\bibfield  {title} {\bibinfo {title} {{GetDist: a Python package
  for analysing Monte Carlo samples}},\ }\href@noop {} {\  (\bibinfo {year}
  {2019})},\ \bibinfo {note} {{\url{https://getdist.readthedocs.io}}},\ \Eprint
  {https://arxiv.org/abs/1910.13970} {arXiv:1910.13970 [astro-ph.IM]}
  \BibitemShut {NoStop}%
\bibitem [{\citenamefont {Ade}\ \emph {et~al.}(2016{\natexlab{b}})\citenamefont
  {Ade} \emph {et~al.}}]{Ade:2015zua}%
  \BibitemOpen
  \bibfield  {author} {\bibinfo {author} {\bibfnamefont {P.~A.~R.}\
  \bibnamefont {Ade}} \emph {et~al.} (\bibinfo {collaboration} {Planck}),\
  }\bibfield  {title} {\bibinfo {title} {{Planck 2015 results. XV.
  Gravitational lensing}},\ }\href
  {https://doi.org/10.1051/0004-6361/201525941} {\bibfield  {journal} {\bibinfo
   {journal} {Astron. Astrophys.}\ }\textbf {\bibinfo {volume} {594}},\
  \bibinfo {pages} {A15} (\bibinfo {year} {2016}{\natexlab{b}})},\ \Eprint
  {https://arxiv.org/abs/1502.01591} {arXiv:1502.01591 [astro-ph.CO]}
  \BibitemShut {NoStop}%
\bibitem [{\citenamefont {Cooke}\ \emph {et~al.}(2018)\citenamefont {Cooke},
  \citenamefont {Pettini},\ and\ \citenamefont {Steidel}}]{Cooke:2017cwo}%
  \BibitemOpen
  \bibfield  {author} {\bibinfo {author} {\bibfnamefont {R.~J.}\ \bibnamefont
  {Cooke}}, \bibinfo {author} {\bibfnamefont {M.}~\bibnamefont {Pettini}},\
  and\ \bibinfo {author} {\bibfnamefont {C.~C.}\ \bibnamefont {Steidel}},\
  }\bibfield  {title} {\bibinfo {title} {{One Percent Determination of the
  Primordial Deuterium Abundance}},\ }\href
  {https://doi.org/10.3847/1538-4357/aaab53} {\bibfield  {journal} {\bibinfo
  {journal} {\apj}\ }\textbf {\bibinfo {volume} {855}},\ \bibinfo {pages} {102}
  (\bibinfo {year} {2018})},\ \Eprint {https://arxiv.org/abs/1710.11129}
  {arXiv:1710.11129 [astro-ph.CO]} \BibitemShut {NoStop}%
\bibitem [{\citenamefont {Mossa}\ \emph {et~al.}(2020)\citenamefont {Mossa}
  \emph {et~al.}}]{Mossa:2020gjc}%
  \BibitemOpen
  \bibfield  {author} {\bibinfo {author} {\bibfnamefont {V.}~\bibnamefont
  {Mossa}} \emph {et~al.},\ }\bibfield  {title} {\bibinfo {title} {{The baryon
  density of the Universe from an improved rate of deuterium burning}},\ }\href
  {https://doi.org/10.1038/s41586-020-2878-4} {\bibfield  {journal} {\bibinfo
  {journal} {Nature}\ }\textbf {\bibinfo {volume} {587}},\ \bibinfo {pages}
  {210} (\bibinfo {year} {2020})}\BibitemShut {NoStop}%
\bibitem [{\citenamefont {Beutler}\ \emph {et~al.}(2011)\citenamefont
  {Beutler}, \citenamefont {Blake}, \citenamefont {Colless}, \citenamefont
  {Jones}, \citenamefont {Staveley-Smith}, \citenamefont {Campbell},
  \citenamefont {Parker}, \citenamefont {Saunders},\ and\ \citenamefont
  {Watson}}]{Beutler:2011hx}%
  \BibitemOpen
  \bibfield  {author} {\bibinfo {author} {\bibfnamefont {F.}~\bibnamefont
  {Beutler}}, \bibinfo {author} {\bibfnamefont {C.}~\bibnamefont {Blake}},
  \bibinfo {author} {\bibfnamefont {M.}~\bibnamefont {Colless}}, \bibinfo
  {author} {\bibfnamefont {D.~H.}\ \bibnamefont {Jones}}, \bibinfo {author}
  {\bibfnamefont {L.}~\bibnamefont {Staveley-Smith}}, \bibinfo {author}
  {\bibfnamefont {L.}~\bibnamefont {Campbell}}, \bibinfo {author}
  {\bibfnamefont {Q.}~\bibnamefont {Parker}}, \bibinfo {author} {\bibfnamefont
  {W.}~\bibnamefont {Saunders}},\ and\ \bibinfo {author} {\bibfnamefont
  {F.}~\bibnamefont {Watson}},\ }\bibfield  {title} {\bibinfo {title} {{The 6dF
  Galaxy Survey: Baryon Acoustic Oscillations and the Local Hubble Constant}},\
  }\href {https://doi.org/10.1111/j.1365-2966.2011.19250.x} {\bibfield
  {journal} {\bibinfo  {journal} {Mon. Not. Roy. Astron. Soc.}\ }\textbf
  {\bibinfo {volume} {416}},\ \bibinfo {pages} {3017} (\bibinfo {year}
  {2011})},\ \Eprint {https://arxiv.org/abs/1106.3366} {arXiv:1106.3366
  [astro-ph.CO]} \BibitemShut {NoStop}%
\bibitem [{\citenamefont {Ross}\ \emph {et~al.}(2015)\citenamefont {Ross},
  \citenamefont {Samushia}, \citenamefont {Howlett}, \citenamefont {Percival},
  \citenamefont {Burden},\ and\ \citenamefont {Manera}}]{Ross:2014qpa}%
  \BibitemOpen
  \bibfield  {author} {\bibinfo {author} {\bibfnamefont {A.~J.}\ \bibnamefont
  {Ross}}, \bibinfo {author} {\bibfnamefont {L.}~\bibnamefont {Samushia}},
  \bibinfo {author} {\bibfnamefont {C.}~\bibnamefont {Howlett}}, \bibinfo
  {author} {\bibfnamefont {W.~J.}\ \bibnamefont {Percival}}, \bibinfo {author}
  {\bibfnamefont {A.}~\bibnamefont {Burden}},\ and\ \bibinfo {author}
  {\bibfnamefont {M.}~\bibnamefont {Manera}},\ }\bibfield  {title} {\bibinfo
  {title} {{The clustering of the SDSS DR7 main Galaxy sample \textendash{} I.
  A 4 per cent distance measure at $z = 0.15$}},\ }\href
  {https://doi.org/10.1093/mnras/stv154} {\bibfield  {journal} {\bibinfo
  {journal} {Mon. Not. Roy. Astron. Soc.}\ }\textbf {\bibinfo {volume} {449}},\
  \bibinfo {pages} {835} (\bibinfo {year} {2015})},\ \Eprint
  {https://arxiv.org/abs/1409.3242} {arXiv:1409.3242 [astro-ph.CO]}
  \BibitemShut {NoStop}%
\bibitem [{\citenamefont {Alam}\ \emph {et~al.}(2021)\citenamefont {Alam} \emph
  {et~al.}}]{eBOSS:2020yzd}%
  \BibitemOpen
  \bibfield  {author} {\bibinfo {author} {\bibfnamefont {S.}~\bibnamefont
  {Alam}} \emph {et~al.} (\bibinfo {collaboration} {eBOSS}),\ }\bibfield
  {title} {\bibinfo {title} {{Completed SDSS-IV extended Baryon Oscillation
  Spectroscopic Survey: Cosmological implications from two decades of
  spectroscopic surveys at the Apache Point Observatory}},\ }\href
  {https://doi.org/10.1103/PhysRevD.103.083533} {\bibfield  {journal} {\bibinfo
   {journal} {Phys. Rev. D}\ }\textbf {\bibinfo {volume} {103}},\ \bibinfo
  {pages} {083533} (\bibinfo {year} {2021})},\ \Eprint
  {https://arxiv.org/abs/2007.08991} {arXiv:2007.08991 [astro-ph.CO]}
  \BibitemShut {NoStop}%
\bibitem [{\citenamefont {Rosenberg}\ \emph {et~al.}(2022)\citenamefont
  {Rosenberg}, \citenamefont {Gratton},\ and\ \citenamefont
  {Efstathiou}}]{Rosenberg:2022sdy}%
  \BibitemOpen
  \bibfield  {author} {\bibinfo {author} {\bibfnamefont {E.}~\bibnamefont
  {Rosenberg}}, \bibinfo {author} {\bibfnamefont {S.}~\bibnamefont {Gratton}},\
  and\ \bibinfo {author} {\bibfnamefont {G.}~\bibnamefont {Efstathiou}},\
  }\bibfield  {title} {\bibinfo {title} {{CMB power spectra and cosmological
  parameters from Planck PR4 with CamSpec}},\ }\href@noop {} {\  (\bibinfo
  {year} {2022})},\ \Eprint {https://arxiv.org/abs/2205.10869}
  {arXiv:2205.10869 [astro-ph.CO]} \BibitemShut {NoStop}%
\bibitem [{\citenamefont {Efstathiou}\ and\ \citenamefont
  {Gratton}(2019)}]{Efstathiou:2019mdh}%
  \BibitemOpen
  \bibfield  {author} {\bibinfo {author} {\bibfnamefont {G.}~\bibnamefont
  {Efstathiou}}\ and\ \bibinfo {author} {\bibfnamefont {S.}~\bibnamefont
  {Gratton}},\ }\bibfield  {title} {\bibinfo {title} {{A Detailed Description
  of the CamSpec Likelihood Pipeline and a Reanalysis of the Planck High
  Frequency Maps}}\ }\href {https://doi.org/10.21105/astro.1910.00483}
  {10.21105/astro.1910.00483} (\bibinfo {year} {2019}),\ \Eprint
  {https://arxiv.org/abs/1910.00483} {arXiv:1910.00483 [astro-ph.CO]}
  \BibitemShut {NoStop}%
\bibitem [{\citenamefont {Gratton}\ and\ \citenamefont
  {Challinor}(2020)}]{Gratton:2019fru}%
  \BibitemOpen
  \bibfield  {author} {\bibinfo {author} {\bibfnamefont {S.}~\bibnamefont
  {Gratton}}\ and\ \bibinfo {author} {\bibfnamefont {A.}~\bibnamefont
  {Challinor}},\ }\bibfield  {title} {\bibinfo {title} {{Understanding
  parameter differences between analyses employing nested data subsets}},\
  }\href {https://doi.org/10.1093/mnras/staa2996} {\bibfield  {journal}
  {\bibinfo  {journal} {Mon. Not. Roy. Astron. Soc.}\ }\textbf {\bibinfo
  {volume} {499}},\ \bibinfo {pages} {3410} (\bibinfo {year} {2020})},\ \Eprint
  {https://arxiv.org/abs/1911.07754} {arXiv:1911.07754 [astro-ph.IM]}
  \BibitemShut {NoStop}%
\bibitem [{\citenamefont {Marzouk}\ \emph {et~al.}(2022)\citenamefont
  {Marzouk}, \citenamefont {Lewis},\ and\ \citenamefont
  {Carron}}]{Marzouk:2022utf}%
  \BibitemOpen
  \bibfield  {author} {\bibinfo {author} {\bibfnamefont {K.}~\bibnamefont
  {Marzouk}}, \bibinfo {author} {\bibfnamefont {A.}~\bibnamefont {Lewis}},\
  and\ \bibinfo {author} {\bibfnamefont {J.}~\bibnamefont {Carron}},\
  }\bibfield  {title} {\bibinfo {title} {{Constraints on $\tau_\mathrm{NL}$
  from Planck temperature and polarization}},\ }\href@noop {} {\  (\bibinfo
  {year} {2022})},\ \Eprint {https://arxiv.org/abs/2205.14408}
  {arXiv:2205.14408 [astro-ph.CO]} \BibitemShut {NoStop}%
\bibitem [{\citenamefont {Zonca}\ \emph {et~al.}(2019)\citenamefont {Zonca},
  \citenamefont {Singer}, \citenamefont {Lenz}, \citenamefont {Reinecke},
  \citenamefont {Rosset}, \citenamefont {Hivon},\ and\ \citenamefont
  {Gorski}}]{Zonca2019}%
  \BibitemOpen
  \bibfield  {author} {\bibinfo {author} {\bibfnamefont {A.}~\bibnamefont
  {Zonca}}, \bibinfo {author} {\bibfnamefont {L.}~\bibnamefont {Singer}},
  \bibinfo {author} {\bibfnamefont {D.}~\bibnamefont {Lenz}}, \bibinfo {author}
  {\bibfnamefont {M.}~\bibnamefont {Reinecke}}, \bibinfo {author}
  {\bibfnamefont {C.}~\bibnamefont {Rosset}}, \bibinfo {author} {\bibfnamefont
  {E.}~\bibnamefont {Hivon}},\ and\ \bibinfo {author} {\bibfnamefont
  {K.}~\bibnamefont {Gorski}},\ }\bibfield  {title} {\bibinfo {title} {healpy:
  equal area pixelization and spherical harmonics transforms for data on the
  sphere in python},\ }\href {https://doi.org/10.21105/joss.01298} {\bibfield
  {journal} {\bibinfo  {journal} {Journal of Open Source Software}\ }\textbf
  {\bibinfo {volume} {4}},\ \bibinfo {pages} {1298} (\bibinfo {year}
  {2019})}\BibitemShut {NoStop}%
\bibitem [{\citenamefont {G\'orski}\ \emph {et~al.}(2005)\citenamefont
  {G\'orski}, \citenamefont {Hivon}, \citenamefont {Banday}, \citenamefont
  {Wandelt}, \citenamefont {Hansen}, \citenamefont {Reinecke},\ and\
  \citenamefont {Bartelman}}]{Gorski:2004by}%
  \BibitemOpen
  \bibfield  {author} {\bibinfo {author} {\bibfnamefont {K.~M.}\ \bibnamefont
  {G\'orski}}, \bibinfo {author} {\bibfnamefont {E.}~\bibnamefont {Hivon}},
  \bibinfo {author} {\bibfnamefont {A.~J.}\ \bibnamefont {Banday}}, \bibinfo
  {author} {\bibfnamefont {B.~D.}\ \bibnamefont {Wandelt}}, \bibinfo {author}
  {\bibfnamefont {F.~K.}\ \bibnamefont {Hansen}}, \bibinfo {author}
  {\bibfnamefont {M.}~\bibnamefont {Reinecke}},\ and\ \bibinfo {author}
  {\bibfnamefont {M.}~\bibnamefont {Bartelman}},\ }\bibfield  {title} {\bibinfo
  {title} {{HEALPix - A Framework for high resolution discretization, and fast
  analysis of data distributed on the sphere}},\ }\href
  {https://doi.org/10.1086/427976} {\bibfield  {journal} {\bibinfo  {journal}
  {Astrophys. J.}\ }\textbf {\bibinfo {volume} {622}},\ \bibinfo {pages} {759}
  (\bibinfo {year} {2005})},\ \Eprint {https://arxiv.org/abs/astro-ph/0409513}
  {arXiv:astro-ph/0409513} \BibitemShut {NoStop}%
\end{thebibliography}%

\end{document}